\documentclass[twocolumn,aps,nofootinbib,floatfix]{revtex4}
\usepackage{graphicx,subfigure}
\usepackage{multirow}
\usepackage[dvips]{epsfig}
\usepackage{amsmath,amssymb,mathrsfs,wasysym,latexsym}
\usepackage{cases,color}
\usepackage{feynmp}
\newcommand{\schi}{\tilde{\chi}^0_i}
\newcommand{\stau}{\tilde{\tau}}
\newcommand{\phase}{i \phi_{\stau}}
\newcommand{\dLips}{{\rm d}\mathscr{L}\!\textsl{ips}}
\newcommand{\ACP}{{\mathcal{A}_\tau^{\rm CP}}}
\renewcommand{\Im}{\mathfrak{Im}}
\renewcommand{\Re}{\mathfrak{Re}}

\begin{document}
\title{Testing the CP-violating MSSM in stau decays at the LHC and ILC}
\author{Herbi~Dreiner$^1$, Olaf~Kittel$^2$, Suchita~Kulkarni$^1$,
  Anja~Marold$^1$ } 

\affiliation{
             $^1$ Bethe Center for Theoretical Physics \& Physikalisches
                  Institut, Universit\"at Bonn, D-53115 Bonn, Germany\\ 
             $^2$ Departamento de F\'isica Te\'orica y del Cosmos and CAFPE,
                  Universidad de Granada, E-18071 Granada, Spain \&
                  II. Institut f\"ur Theoretische Physik, Universit\"at Hamburg,
                  D-22761 Hamburg, Germany}
\begin{abstract}
We study CP violation in the two-body decay of a scalar tau into a
neutralino and a tau, which should be probed at the LHC and ILC.
From the normal tau polarization, a CP asymmetry is defined which is
sensitive to the CP phases of the trilinear scalar coupling parameter
$A_\tau$, the gaugino mass parameter $M_1$, and the higgsino mass parameter
$\mu$ in the stau-neutralino sector of the Minimal Supersymmetric Standard
Model. Asymmetries of more than $70\%$ are obtained in scenarios with strong
stau mixing. As a result, detectable CP asymmetries in stau decays at
the LHC are found, motivating further  detailed experimental studies
for probing the SUSY CP phases.  
\end{abstract}
\maketitle

\section{Introduction}\label{sec:intro}
The surplus of matter over anti-matter within the universe can only be
explained with a thorough understanding of CP violation. The CP phase
in the quark mixing matrix of the Standard Model, which has been
confirmed by B-meson experiments~\cite{Belle}, is not sufficient  to
understand the baryon asymmetry of the Universe~\cite{Sakharov}. However, the  
Minimal Supersymmetric Standard Model (MSSM)~\cite{Haber:1984rc} provides new
physical phases that are manifestly CP-sensitive. After absorbing
non-physical phases, we chose the complex parameters to be the higgsino mass  
parameter $\mu$, the U(1), and SU(3) gaugino mass parameters $M_1$, and $M_3$,
and the trilinear scalar coupling parameters $A_f$ of the third generation
sfermions $(f=b,t,\tau)$. The corresponding phases violate CP and
are generally constrained by experimental bounds on electric
dipole moments (EDMs)~\cite{Li:2010ax}. However, these restrictions are
strongly model
dependent~\cite{cancellations3,cancellations1,cancellations2}, such
that additional measurements outside the low energy EDM sector are
required.\\

Many CP observables have been proposed and studied in order to measure
CP violation. Total cross sections~\cite{totalsigma}, masses~\cite{masses},
and branching ratios~\cite{BRs}, are CP-even quantities. For a direct
evidence of CP violation, however,  CP-odd (T-odd) observables are
required. Examples are rate asymmetries of either branching
ratios~\cite{rateasymBRs}, cross sections~\cite{rateasymsigma}, or angular
distributions~\cite{angulardistrib}. Since these rate asymmetries
require the presence of absorptive phases, they are typically small,
of the order of $<10\%$, if they are not resonantly enhanced~\cite{res}.
Larger CP-odd observables which already appear at tree-level are desirable. 
These are T-odd triple products of momenta and/or spins,
from which CP-odd asymmetries  can be constructed. Such triple product
asymmetries are highly CP-sensitive, and have been intensively studied
both at lepton and hadron colliders~\cite{tripleproducts,CPreview}.\\

Third generation sfermions have a rich phenomenology at high energy
colliders like the LHC~\cite{LHC} or ILC~\cite{ILC} due to a sizable 
mixing of left and right states. In addition, the CP phases of the
trilinear coupling parameters $A_f$ are rather unconstrained by the
EDMs~\cite{Semertzidis:2004uu,cancellations3,Choi:2004rf}. The phases
of $A_{b}$ and $A_{t}$ have been studied in
stop~\cite{Bartl:2004jr,Ellis:2008hq,Deppisch:2009nj,MoortgatPick:2009jy}
and  sbottom~\cite{Bartl:2006hh,Deppisch:2010nc} decays, respectively.
Since these are decays of a scalar particle, the spin-spin correlations
have to be taken into account. The triple product asymmetries can then be
up to $40\%$, for sizable squark mixing. Similarly for probing the
CP-violating phase of $A_{\tau}$ in the stau vertex,
$\tilde\tau$-$\tilde\chi^0$-$\tau$, it is essential to include the tau spin.
Only then is there a sensitivity to the phase of
$A_{\tau}$~\cite{Bartl:2003gr,Bartl:2003ck}.  If the spin of the tau
is summed over, this crucial information is lost. Triple product
asymmetries including the tau polarization have been studied in  
neutralino decays $\tilde \chi_i^0\to \tilde\tau \tau$~\cite{Bartl:2003gr}, 
and also in chargino decays $\tilde \chi_i^\pm\to \tilde\nu_\tau
\tau^\pm$~\cite{MKD}. It was shown that the normal tau polarization
itself is CP-sensitive,  and that the asymmetries are large and of the
order of $60\%$ to $70\%$.\\

\begin{figure}[t!]
\centering
\includegraphics[angle =0,width=0.74\columnwidth]{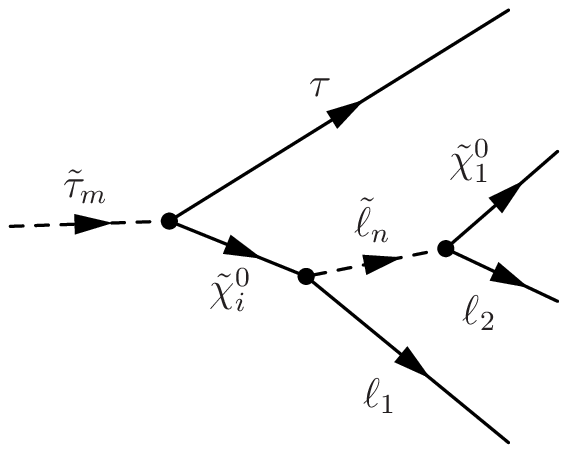}
\caption{Schematic picture of stau decay.}
\label{Feyn}
\end{figure}

We are thus motivated to study CP violation, including the tau polarization, 
in the two-body decay of a stau
\begin{equation}
\tilde\tau_m      \to   \tau   +\schi, \quad m=1,2, \quad
i=2,3,4, \label{eq:staudec} 
\end{equation}
followed by the subsequent chain of two-body decays 
\begin{subequations}
\label{eq:decaychain}
\begin{align}
\label{eq:decayChi}
        \tilde\chi^0_i &\to \ell_1         + \tilde\ell_{n}; \\
\label{eq:decaySel}
        \tilde\ell_{n} &\to \tilde\chi^0_1 + \ell_2;     \quad n=L,R,
        \quad  \ell= e,\mu.
\end{align} 
\end{subequations}
See Fig.~\ref{Feyn} for a schematic picture of the entire stau decay.
This process is kinematically open for a mass hierarchy 
\begin{eqnarray}
m_{\tilde \tau} &>& m_{\tilde \chi_i^0} \;> \;m_{\tilde e}\,= \, m_{\tilde \mu},
\end{eqnarray}
where the staus are heavier than the smuons and selectrons.
We thus work in MSSM scenarios with heavier stau soft SUSY 
breaking parameters 
\begin{eqnarray}
M_{ \tilde E_\tau} &>& M_{\tilde E_e} = M_{\tilde E_\mu} \\
M_{ \tilde L_\tau} &>& M_{\tilde L_e} = M_{\tilde L_\mu}.
\end{eqnarray}
We show that the normal tau polarization, 
with respect to the plane spanned by the
$\tau$  and  $\ell_1$ momentum, is a 
triple product asymmetry which is sensitive to the phases
of $A_\tau$, $M_1$, and $\mu$ in the stau-neutralino sector.
For nearly degenerate stau masses, 
$M_{ \tilde E_\tau} \approx M_{ \tilde L_\tau},$ 
a strong stau mixing is obtained which results
 in tau polarization asymmetries of more than $70\%$. 
This should be measurable at 
colliders\footnote{Note that we do not include the  tau decay in our
  calculations. However, some of the decay products of the tau
  have to be reconstructed in order to measure the tau spin.
  The main goal of our work is to motivate such an experimental study,
  to address the feasibility of measuring the CP phases at the LHC or ILC.
}.
Since the stau is a scalar particle, its particular production 
does not contribute to CP-sensitive spin-spin correlations,
and can thus be considered separately.
This allows a collider-independent study, where we 
only discuss the boost dependence of the CP asymmetries.\\

The paper is organized as follows. In Section~\ref{sec:formalism} we review
stau mixing and the stau-neutralino Lagrangian with complex couplings. 
We calculate the amplitude squared for the entire stau decay in the
spin-density  matrix formalism~\cite{Haber:1994pe}. We  construct the
CP asymmetry from the normal tau polarization, and discuss its MSSM
parameter dependence, as well its boost dependence for colliders like
the ILC and LHC. In Section~\ref{sec:results}, we numerically study
the phase and  parameter dependence of the asymmetry, and the stau and
neutralino branching ratios. We comment on the impact of the
${\tilde\tau}_2$ decay in  scenarios  with nearly degenerate stau masses. 
We summarize and conclude in Section~\ref{sec:summary}. The Appendices
contain the definitions of momenta and spin vectors,  the analytical
expressions for the stau decay amplitudes  in the spin-density matrix
formalism,  and formulae for the stau decay widths.

\section{Formalism}\label{sec:formalism}

\subsection{Stau mixing}\label{sub:mixing}
In the complex MSSM, the stau mixing matrix in the $(\stau_L,\stau_R)$-basis
is~\cite{Haber:1984rc,Bartl:2002uy} 
\begin{equation}
\mathcal{M}_{\tilde\tau} = 
\left( \begin{array}{cc}
m^2_{\stau_L} & e^{- \phase} m_{\tau}|\Lambda_{\stau}| \\
\\
e^{\phase} m_{\tau} |\Lambda_{\stau}| & m^2_{\stau_R} \\
\end{array} \right).
\end{equation} 

CP violation is parameterized by the physical phase
\begin{eqnarray} \label{eq:phi_stau}
\phi_{\stau} & = & {\rm arg} [\Lambda_{\stau}], \\
\Lambda_{\stau} & = & A_\tau - \mu^* \cot{\beta},
\end{eqnarray}
with the complex trilinear scalar coupling parameter $A_\tau$,
the complex higgisino mass parameter $\mu$, and  $\tan\beta = v_1/v_2$,
 the ratio of the vacuum expectation values of the two neutral Higgs fields.
The  left and right stau masses are
\begin{eqnarray}
 m^2_{\stau_L} & = & M^2_{{\tilde{L}}_\tau} + (-\frac{1}{2} + \sin^2\theta_w)
                      m^2_Z \cos(2\beta) + m^2_\tau , 
\label{eq:left_stau}\\
 m^2_{\stau_R} & = & M^2_{{\tilde{E}}_\tau} -  \sin^2\theta_w  m^2_Z \cos(2\beta)
                     + m^2_\tau,
\label{eq:right_stau}
\end{eqnarray}
with the real soft SUSY breaking parameters
$M^2_{{\tilde{L}_\tau},{\tilde{E}}_\tau }$, 
the electroweak mixing angle $\theta_w$, and the masses of the $Z$ boson
$m_Z$,  
and of the tau lepton, $m_\tau$. \\

In the mass basis, the stop eigenstates are~\cite{Haber:1984rc,Bartl:2002uy}
\begin{equation}
\left( \begin{array}{c} \stau_1 \\ 
                        \stau_2  \end{array} \right)
 = \mathcal{R}^{\stau} \left( \begin{array}{c}
                                \stau_L \\
                                \stau_R  \end{array} \right),
\end{equation}
with the diagonalization matrix  
\begin{equation}
\mathcal{R}^{\stau} = \left( \begin{array}{cc} e^{\phase}\cos\theta_{\stau} & 
\sin\theta_{\stau} \\
 -\sin\theta_{\stau} & e^{-\phase}\cos\theta_{\stau} \end{array} \right),
\label{eq:rstau}
\end{equation}
and the stau mixing angle
\begin{eqnarray}
\cos\theta_{\stau} &=& \frac{\displaystyle -m_{\tau}
  |\Lambda_{\stau}|}{\displaystyle \sqrt{m_{\tau}^2|\Lambda_{\stau}^2|+ 
\left( m_{\stau_1}^2-m_{\stau_2}^2\right)^2}} ,
\label{eq:cos_theta_stau}\\
\sin\theta_{\stau} &=& \frac{\displaystyle m_{\stau_L}^2 -
  m_{\stau_1}^2}{\displaystyle \sqrt{m_{\tau}^2|\Lambda_{\stau}^2|+ 
\left( m_{\stau_1}^2-m_{\stau_2}^2\right)^2}}.
\label{eq:sin_theta_stau}
\end{eqnarray} 

The stau mass eigenvalues are  
\begin{eqnarray}
m^2_{\stau_{1,2}} &=& \frac{1}{2} \bigg[ \left( m_{\stau_L}^2 + m_{\stau_R}^2
\right)  
\mp 
\nonumber \\
&&
\phantom{ \frac{1}{2}}
\sqrt {\left(
    m_{\stau_L}^2-m_{\stau_R}^2\right)^2+4m_{\stau}^2|\Lambda_{\stau}|^2}
~\bigg].  
\end{eqnarray}

\subsection{Lagrangian and complex couplings}\label{sub:lagrang}
The relevant Lagrangian terms for the stau decay
$\stau_m \to \tau\tilde\chi_i^0 $
are~\cite{Haber:1984rc,Bartl:2002uy}
\begin{eqnarray}
{\mathscr L}_{\tau \stau \tilde\chi^0}
= g\,\bar \tau\,(a_{mi}^{\stau}\,P_R + b_{mi}^{\stau}\,P_L)
            \,\tilde\chi_i^0\, \stau_m
+ {\rm h.c.} ,
\label{eq:stauLag}
\end{eqnarray}
with $P_{L, R}=(1\mp \gamma_5)/2$, and the weak coupling constant
$g=e/\sin\theta_w$, $e>0$.
The couplings are defined as~\cite{Bartl:2002uy}
\begin{equation}
a_{mi}^{\stau} \equiv \sum^2_{n=1}\, (\mathcal{R}^{\stau}_{mn})^{\ast}\,
                    {\cal A}_{in}^\tau, \quad
b_{mi}^{\stau} \equiv \sum^2_{n=1}\, (\mathcal{R}^{\stau}_{mn})^{\ast}\,
                    {\cal B}_{in}^{\tau}. 
\label{eq:abstop}
\end{equation}
The stau diagonalization matrix 
$\mathcal{R}^{\tilde t}$ is given in  Eq.~(\ref{eq:decaychain}), and
\begin{equation}
{\cal A}_i^\tau \equiv
  \left(\begin{array}{ccc}
   f_{\tau i}^L \\[2mm]
   h_{\tau i}^R \end{array}\right), \qquad
  {\cal B}_i^\tau \equiv
 \left(\begin{array}{ccc}
    h_{\tau i}^L\\[2mm]
    f_{\tau i}^R\end{array}\right).
\label{eq:AB}
\end{equation}
In the photino, zino, higgsino basis
($\tilde{\gamma},\tilde{Z}, \tilde{H}^0_a, \tilde{H}^0_b$),
we have
\begin{eqnarray}
f_{\tau i}^L &=& \sqrt{2}\bigg[\frac{1}{\cos
        \theta_w}\left(\frac{1}{2}-\sin^2\theta_w\right)N_{i2}+
        \sin \theta_w N_{i1}\bigg], \qquad
\label{eq:flt}\\[2mm]
 f_{\tau i}^R &=& \sqrt{2} \sin \theta_w
               \left(\tan\theta_w N_{i2}^*-N_{i1}^*\right),
\label{eq:frt}\\[2mm]
 h_{\tau i}^L &=& (h_{\tau i}^R)^{\ast} = 
            -Y_\tau( N_{i3}^\ast\cos\beta+N_{i4}^\ast\sin\beta),
\label{eq:hrlt}\\[2mm]
      Y_\tau  &=& \frac{m_\tau}{\sqrt{2}\,m_W \cos\beta}, 
\label{eq:yt}
\end{eqnarray}
with $m_W$ the mass of the $W$ boson,
and $N$ the  complex, unitary $4\times 4$ matrix that diagonalizes 
the neutralino mass matrix~\cite{Haber:1984rc}
\begin{equation}
        N^* \cdot {\mathcal M}_{\tilde\chi^0} \cdot N^{\dagger} =
        {\rm diag}(m_{\tilde\chi^0_1},\dots,m_{\tilde\chi^0_4}).
\label{eq:neutn}
\end{equation}
The interaction Lagrangian relevant for the neu\-tra\-lino decay
$\tilde\chi_i^0 \to \tilde\ell_{R,L}^\pm \ell^\mp$, for $\ell = e,\mu$
is~\cite{Haber:1984rc}
\begin{eqnarray}
      {\mathscr L}_{\ell \tilde \ell \tilde\chi^0} & = & 
             g \bar\ell f_{\ell i}^L  P_R \tilde\chi_i^0  \tilde \ell_L 
     +       g \bar\ell f_{\ell i}^R  P_L \tilde\chi_i^0  \tilde \ell_R
     +       \mbox{h.c.},
\label{eq:slechie}
\end{eqnarray}
with the couplings $f_{\ell i}^{L,R}$ 
given in Eqs.~\eqref{eq:flt} and~\eqref{eq:frt}.

\subsection{Tau spin density matrix}\label{sub:matrix}
The unnormalized, $2\times 2$, hermitian, $\tau$ spin density matrix 
for stau decay, Eqs.~\eqref{eq:staudec} and~\eqref{eq:decaychain}, reads 
\begin{equation} 
  \rho^{\lambda_\tau\lambda^\prime_\tau}\equiv  
  \int\left(|\mathcal{M}|^2\right)^{\lambda_\tau\lambda^\prime_\tau}{\dLips}, 
\label{rho} 
\end{equation} 
with the amplitude $\mathcal{M}$, and the Lorentz invariant phase
space element ${\dLips}$, for details see
Appendix~\ref{sec:phasespace}.  The $\tau$ helicities are denoted by
$\lambda_\tau$ and $\lambda^\prime_\tau$. In the spin density matrix
formalism~\cite{Haber:1994pe}, the amplitude squared is given by
\begin{eqnarray}
\lefteqn{
\left(|\mathcal{M}|^2\right)^{\lambda_\tau\lambda^\prime_\tau}
= 
|\Delta(\schi)|^2|\Delta(\tilde\ell)|^2\times
}
\nonumber\\[2mm]
&&\sum_{\lambda_i \lambda_i^\prime}
\rho_D(\tilde\tau)^{\lambda_\tau\lambda_\tau^\prime}
_{\lambda_i\lambda_i^\prime}
\;\rho_{D_1}(\schi)^{\lambda_i^\prime\lambda_i}
\;D_2(\tilde\ell),
\label{rhoampli}
\end{eqnarray}
with the neutralino helicities $\lambda_i,\, \lambda_i^\prime$.
The amplitude squared decomposes into the remnants of the propagators
\begin{equation}
\Delta(j)=\dfrac{i}{s_j-m_j^2+im_j\Gamma_j},
\end{equation}
with mass $m_j$, and width $\Gamma_j$ of particle $j=\schi$ or $\tilde\ell$,
and the unnormalized spin density matrices for stau decay $\rho_D(\tilde\tau)$,
and neutralino decay $\rho_{D_1}(\schi)$. The decay matrix of the spinless
slepton is a factor since the polarizations of the final lepton and
LSP are not accessible. The corresponding amplitude is denoted by    
$D_2(\tilde\ell)$.  Defining  a set of spin basis vectors $s_\tau^a$ for the 
tau, see Eqs.~\eqref{eq:spintau} in Appendix~\ref{sec:momenta}, and
$s_{\schi}^b$ for the neutralino \cite{Kittel:2004rp}, the spin
density matrices can be expanded in terms of the Pauli matrices~$\sigma$
\begin{eqnarray}
\rho_{\rm D}(\tilde\tau)^{\lambda_\tau\lambda_\tau^\prime}
_{\lambda_i\lambda_i^\prime}=
{\rm D}\,\delta^{\lambda_\tau\lambda_\tau^\prime}
         \delta_{\lambda_i\lambda_i^\prime}
 + \Sigma_{\rm D}^a\,(\sigma^a)^{\lambda_\tau\lambda_\tau^\prime}
                      \delta_{\lambda_i\lambda_i^\prime}
                   +
\nonumber\\[2mm]
\Sigma_{\rm D}^b\,\delta^{\lambda_\tau\lambda_\tau^\prime}
                   (\sigma^b)_{\lambda_i\lambda_i^\prime}
 +
\Sigma_{\rm D}^{ab}\,(\sigma^a)^{\lambda_\tau\lambda_\tau^\prime}
                       (\sigma^b)_{\lambda_i\lambda_i^\prime},
\label{rhoD}\\[2mm]
\rho_{{\rm D}_1}({\tilde\chi}_i^0)^{\lambda_i^\prime\lambda_i}=
{\rm D}_1\,\delta^{\lambda_i^\prime\lambda_i}
+\Sigma_{{\rm D}_1}^b\,(\sigma^b)^{\lambda_i^\prime\lambda_i},
\phantom{spacespace}
\label{rhoD1}
\end{eqnarray}
with an implicit sum over $a,b=1,2,3$, respectively.
The real expansion coefficients ${\rm D}$, ${\rm D_1}$, $\Sigma_{\rm
D}^a$, $\Sigma_{\rm D}^b$, $\Sigma_{{\rm D}_1}^b$ and $\Sigma_{\rm
D}^{ab}$ contain the physical information of the process. ${\rm D}$
denotes the unpolarized part of the amplitude for stau decay
${\tilde\tau}_m \to \chi^0_i \tau$ , ${\rm
  D_1}$ denotes the unpolarized part for neutralino decay $\chi^0_i \to
{\tilde\ell}_R \ell_1 $ , respectively. $\Sigma_{\rm D}^a$ gives the
tau polarization,   $\Sigma_{\rm D}^b$, and $\Sigma_{{\rm D}_1}^b$
describe the contributions 
from the neutralino polarization, and $\Sigma_{\rm D}^{ab}$ is the
spin-spin correlation term, which contains the  CP-sensitive parts. We give the
expansion coefficients explicitly in Appendix~\ref{sec:coefficients}.\\

Inserting the density matrices, Eqs.~\eqref{rhoD} and \eqref{rhoD1}, 
into Eq.~\eqref{rhoampli}, we get for the amplitude squared
\begin{eqnarray} 
(|\mathcal{M}|^2)^{\lambda_\tau\lambda_\tau^\prime}&=& 
                2|\Delta({\tilde\chi^0}_i)|^2  |\Delta(\tilde\ell)|^2
\times\nonumber\\
&&                 \Big[~({\rm D}{\rm D}_1 + \Sigma_{\rm
  D}^b\Sigma_{{\rm D}_1}^b)  
                          \delta^{\lambda_\tau\lambda_\tau^\prime} \nonumber\\ 
&&                      +(  \Sigma_{\rm D}^a  {\rm D}_1 
                          + \Sigma_{\rm D}^{ab} \Sigma_{{\rm D}_1}^{b})   
                         (\sigma^a)^{\lambda_\tau\lambda_\tau^\prime}\Big]{\rm
                           D}_2,\quad 
\label{eq:nosum} 
\end{eqnarray}
with an implicit sum over $a,b=1,2,3$.
The amplitude squared $(|\mathcal{M}|^2)^{\lambda_\tau\lambda_\tau^\prime}$
is now decomposed into an unpolarized part (first summand), 
and into the part for the tau polarization (second summand),
in Eq.~\eqref{eq:nosum}.
By using the completeness relations for the 
neutralino spin vectors, Eq.~(\ref{eq:completeness2}),
the products in Eq.~(\ref{eq:nosum}) can be 
written\footnote{The formulas are given for the decay of a negatively 
                 charged stau $\stau_m \to \tau^- \tilde\chi_i^0 $, 
                 followed by
                 $\tilde\chi^0_i \to \ell_1^+ \tilde \ell_R^-$.
                 The signs in parentheses in
                  Eqs.~(\ref{eq:product1}) and (\ref{eq:product2})
                 hold for the charge conjugated stau decay
                 $\stau_m^{\ast} \to \tau^+\tilde\chi_i^0 $;
                 $\tilde\chi^0_i \to \ell_1^+ \tilde \ell_R^-$.
                 In order to obtain the terms for the decay 
                 $\stau_m^{(\ast)} \to \tau^\mp \tilde\chi_i^0 $, 
                 however, followed by the neutralino decay 
                 into a positively charged slepton, 
                 $\tilde\chi^0_i \to \ell_1^- \tilde \ell_R^+$, 
                 one has to reverse the signs of Eqs.~(\ref{eq:product1}) 
                 and (\ref{eq:product2}). This is due to the sign change
                 of $\Sigma^{b}_{{\rm D}_1}$, see Eqs.~(\ref{eq:SigmabD1}).
                 In Appendix~\ref{sec:coefficients}, we also give the terms for
                 the neutralino decay into a left slepton, 
                 $\tilde\chi^0_i \to \ell_1^\pm \tilde \ell_L^\mp$.
                 Note that the term proportional to $m_\tau$ 
                 in Eq.~\eqref{eq:product2}
                 is negligible at high particle energies $E \gg m_\tau$. 
},
\begin{eqnarray}
\Sigma^b_{\rm D}~\Sigma^{b}_{{\rm D}_1} 
&=&  \,^{\;\,+}_{(-)}
  \frac{g^4}{2}\left(|a^{\stau}_{mi}|^2 - |b^{\stau}_{mi}|^2\right)
   |f^{R}_{\ell i}|^2 \times
  \nonumber\\ [2mm] &&
   \left[
      m_{\tilde\chi_i^0}^2(p_\tau\cdot p_{\ell_1})
      -(p_{\tilde\chi_i^0}\cdot p_\tau)(p_{\ell_1}\cdot p_{\tilde\chi_i^0})
  \right], \quad \label{eq:product1}\\ [2mm]
\Sigma^{ab}_{\rm D}~\Sigma^{b}_{{\rm D}_1} &=& 
     \,^{\;\,+}_{(-)}
       \frac{g^4}{2}\left(|a^{\stau}_{mi}|^2 + |b^{\stau}_{mi}|^2\right)
    |f^{R}_{\ell i}|^2 m_\tau 
 \times\nonumber\\[2mm]&&
\left[
    (s^a_\tau \cdot p_{\tilde\chi_i^0})(p_{\tilde\chi_i^0}\cdot p_{\ell_1})
     -m_{\tilde\chi_i^0}^2  (s^a_\tau \cdot p_{\ell_1} )
\right]
\nonumber \\ [2mm]&&
 \,^{\;\,+}_{(-)} g^4 \Re \{ a^{\stau}_{mi} (b^{\stau}_{mi})^\ast \}
  |f^{R}_{\ell i}|^2  m_{\tilde\chi_i^0} 
 \times\nonumber\\&&
\left[
    (p_\tau \cdot p_{\tilde\chi_i^0})(s^a_\tau \cdot p_{\ell_1})
     - (p_\tau \cdot p_{\ell_1})  (s^a_\tau \cdot p_{\tilde\chi_i^0} )
\right]
\nonumber \\ [2mm]&&
 \,^{\;\,+}_{(-)} g^4 
  |f^{R}_{\ell i}|^2  m_{\tilde\chi_i^0} 
\times\nonumber\\[2mm]&&
\Im \{ a^{\stau}_{mi} (b^{\stau}_{mi})^\ast \}
[p_{\stau},~ p_{\ell_1},~  p_{\tau}, ~s^a_\tau].
\label{eq:product2}
\end{eqnarray}
The spin-spin correlation term 
$\Sigma^{ab}_{\rm D}~\Sigma^{b}_{{\rm D}_1}$, Eq.~\eqref{eq:product2},
explicitly depends on the imaginary part
$ \Im \{ a^{\stau}_{mi} (b^{\stau}_{mi})^\ast \} $ 
of the stau-tau-neutralino couplings,
Eq.~\eqref{eq:stauLag}. Thus this term is manifestly CP-sensitive, 
i.e., it depends on the phases 
$\phi_{A_\tau}$, $\phi_{1}$, $\phi_\mu$
of the stau-tau-neutralino sector.
The imaginary part is multiplied by the totally anti-symmetric
 (epsilon) product,
\begin{equation}
 \mathcal{E}^{a}\equiv
[p_{\stau},~ p_{\ell_1},~  p_{\tau}, ~s^a_\tau] \equiv 
  \epsilon_{\mu\nu\rho\sigma} \, 
      p_{\stau}^\mu \,
      p_{\ell_1}^\nu \,
      p_{\tau}^\rho \, 
      s^{a, \sigma}_{\tau},
\label{eq:epsilon}
\end{equation}
with the convention $\epsilon_{0123}=1$.
Since each of the spatial components of the four-momenta $p$, or the
spin vectors $s^{a}_{\tau}$, changes sign under a time transformation,   
$t \to -t$, the  epsilon product $\mathcal{E}^{a}$  is T-odd.
In the stau rest frame, $p_{\stau}^\mu = (m_{\stau}, \mathbf{0})$, 
the epsilon product reduces to the T-odd triple product ${\mathcal T}^a$
\begin{equation}
[p_{\stau},~ p_{\ell_1},~  p_{\tau}, ~s^a_\tau] =
m_{\stau} \; \, ( \mathbf{p}_{\ell_1 } \times \mathbf{p}_{\tau}) 
\cdot   \mathbf{s}^a_{\tau} \equiv
        m_{\stau} \, {\mathcal T}^a.
\label{eq:tripleterm}
\end{equation}
The task in the next section is to define an observable that
projects out from the amplitude squared the part proportional to  
$\mathcal{E}^{a}$ (or ${\mathcal T}^a$), in order to probe the
CP-sensitive coupling combination 
$ \Im \{ a^{\stau}_{mi} (b^{\stau}_{mi})^\ast \}$.

\subsection{Normal tau polarization and CP asymmetry}
                                                    \label{sub:asymmetry}

The $\tau$ polarization 
is given by the expectation value of the Pauli matrices
{\boldmath $\sigma$}$=(\sigma_1,\sigma_2,\sigma_3)$~\cite{Renard:1981de}
\begin{equation} 
 \mbox{\boldmath $ \mathcal P $ }
= \frac{{\rm Tr}\{\rho \mbox{\boldmath$ \sigma$} \}}{{\rm Tr}\{\rho\}}, 
\label{eq:taupol} 
\end{equation} 
with the $\tau$ spin density matrix $\rho$, as given in  Eq.~\eqref{rho}.
In our convention for the polarization vector
{\boldmath $ \mathcal P $ }$ =(\mathcal{P}_1,\mathcal{P}_2,\mathcal{P}_3)$, 
the components $\mathcal{P}_1$ and $\mathcal{P}_3$ are the transverse and
longitudinal  
polarizations in the plane spanned by ${\bf p}_{\ell_1}$ and ${\bf p}_{\tau}$, 
respectively, and  $\mathcal{P}_2$  is the polarization normal to that plane.
See our definition of the tau spin basis vectors $s_\tau^a$ in
Appendix~\ref{sec:momenta}. \\

The normal $\tau$ polarization is equivalently defined as
\begin{eqnarray} 
\mathcal{P}_2 &\equiv&
\frac{N(\uparrow)-N(\downarrow)}{N(\uparrow)+N(\downarrow)},
\label{eq:tau_norm_pol}
\end{eqnarray}
with the number of events $N$ with the 
$\tau$ spin up $(\uparrow)$ or down $(\downarrow)$, with respect
to the quantization axis  ${\bf p}_{\ell_1}\times{\bf p}_{\tau}$,
see Eq.~\eqref{eq:spintau}.
The normal  $\tau$ polarization can  thus also be regarded as an asymmetry
\begin{equation} 
\mathcal{P}_2= \frac{\sigma(\mathcal{T}>0)-\sigma(\mathcal{T}<0)} 
                   {\sigma(\mathcal{T}>0)+\sigma(\mathcal{T}<0)}, 
\label{eq:tripleasymm}
\end{equation} 
of the triple product 
\begin{equation} 
\mathcal{T}=
({\bf p}_{\ell_1} \times  {\bf p}_{\tau})\cdot
{\mbox{\boldmath$ \xi$} }_{\tau}, 
\label{eq:triple} 
\end{equation} 
where {\boldmath$ \xi$}$_{\tau}$ is the direction of the $\tau$ spin vector
for each event. The triple product $\mathcal{T}$ is included in
the spin-spin correlation term 
$\Sigma^{ab}_{\rm D}~\Sigma^{b}_{{\rm D}_1}$, Eq.~\eqref{eq:product2},
{\it cf.} Eq.~\eqref{eq:tripleterm},
and the asymmetry thus probes the term which contains 
the CP-sensitive coupling combination 
$ \Im \{ a^{\stau}_{mi} (b^{\stau}_{mi})^\ast \}$.\\

Since under  naive time reversal, $t\to -t$,  the triple product 
$\mathcal{T}$ changes sign, the tau polarization 
$\mathcal{P}_2$, Eq.~\eqref{eq:tripleasymm}, is T-odd.
Due to CPT invariance~\cite{Luders:1954zz}, $\mathcal{P}_2$ would thus be
CP-odd at tree level. 
In general, $\mathcal{P}_2$ also has contributions from absorptive phases, 
e.g. from intermediate $s$-state resonances or final-state interactions, 
which do not signal CP violation. Although such absorptive contributions are 
a higher order effect, and thus expected to be small, they can be eliminated
in the true CP asymmetry~\cite{Bartl:2003gr}
\begin{equation} 
\ACP = \frac{1}{2}(\mathcal{P}_2 - \bar{\mathcal{P}}_2), 
\label{eq:asym} 
\end{equation} 
where $\bar{\mathcal{P}}_2$ is the normal tau polarization for the charged 
conjugated process 
$\stau_m^{\ast} \to \tau^+\tilde\chi_i^0 $. 
For our analysis at tree level,  where no absorptive phases are present,
we find $\bar{\mathcal{P}}_2=-\mathcal{P}_2$, 
see the sign change in Eqs.~\eqref{eq:product1}  and \eqref{eq:product2},
and thus $\ACP=\mathcal{P}_2$.
We study $\ACP$ in the following, which
is, however, equivalent to  $\mathcal{P}_2$ at tree level.\\

Inserting now the explicit form of the density matrix $\rho$, Eq.~\eqref{rho},
into Eq.~\eqref{eq:taupol}, together with Eq.~\eqref{eq:nosum}, we obtain  the
CP asymmetry     
\begin{equation} 
\ACP=\mathcal{P}_2 
                         =\frac{\int 
       \Sigma^{a=2, b}_{\rm D}~\Sigma^{b}_{{\rm D}_1} \,{\dLips} } 
                               {\int  {\rm D}{\rm D}_1\,{\dLips}},  
\label{P2} 
\end{equation} 
where we have used the narrow width approximation for the propagators
in the phase space element ${\dLips}$, see Eq.~\eqref{NWA}. 
Note that in the denominator of $\ACP$, Eq.~\eqref{P2}, 
the spin correlation terms 
vanish, $\int  \Sigma^b_{\rm D}~\Sigma^{b}_{{\rm D}_1}  \,{\dLips}=0$,
see Eq.~\eqref{eq:product1},
when integrated over phase space. In the numerator  only 
the spin-spin
correlation term  $\Sigma^{a b}_{\rm D}~\Sigma^{b}_{{\rm D}_1}$
for $a=2$ contributes, which contains the T-odd 
epsilon product $\mathcal{E}^{a}$, see Eq.~\eqref{eq:epsilon}.\\

\subsection{Parameter dependence of the CP asymmetry}
                                \label{sub:analytics}

To qualitatively understand the dependence of the asymmetry 
$\ACP$, Eq.~\eqref{P2},  on the MSSM parameters, 
we study in some detail its dependence 
on the $\stau_m$-$\tau$-$\tilde\chi_i^0$ couplings, 
$a^{\stau}_{mi}$ and $ b^{\stau}_{mi} $, see Eq.~\eqref{eq:lstau}. 
From the explicit form of the decay terms 
$ \Sigma^b_{\rm D}~\Sigma^{b}_{{\rm D}_1} $
Eq.~\eqref{eq:product1}, and
${\rm D}$, ${\rm D}_1$, Eqs.~\eqref{eq:D}, \eqref{eq:D1},
respectively, we find that the asymmetry 
\begin{eqnarray}
\ACP&=&\eta_{mi} \;
           \frac{m_{{\tilde\chi^0}_i}\int  
 [p_{\stau},~ p_{\ell_1},~  p_{\tau}, ~s^{a=2}_\tau]
                         \,\dLips}
                { (p_{{\tilde\chi}_i^0}\cdot p_{\tau}) 
                  (p_{{\tilde\chi}_i^0}\cdot p_{\ell_1})\,\int\dLips},
\label{finalmathcalA}
\end{eqnarray}
with $ (p_{{\tilde\chi}_i^0}\cdot p_{\tau})= 
(m_{\stau}^2 -m_{\tilde\chi_i^0}^2)/2$, 
and
$ (p_{\tilde\chi^0_i}\cdot p_{\ell_1})= 
(m_{\tilde\chi_i^0}^2-m_{\tilde\ell}^2)/2$,
is proportional to the decay coupling factor 
\begin{equation}
 \eta_{mi}=
\frac{  \Im \{ a^{\stau}_{mi} (b^{\stau}_{mi})^\ast \} }
          {\frac{1}{2}(| a^{\stau}_{mi}|^2 + | b^{\stau}_{mi} |^2)},
\label{etafactor}
\end{equation}
with $\eta_{mi} \in [-1,1]$.
We thus expect maximal asymmetries for equal moduli of 
left and right couplings, $|a^{\stau}_{mi}| \approx |b^{\stau}_{mi}|$,
which have a phase difference of about $\pi/2$,
where the coupling factor can be  maximal $\eta_{mi}=\pm1$, see
Eq.~\eqref{etafactor}.\\

To study the dependence of $\eta$ on the CP phase $\phi_{\tilde\tau}$ 
of the stau sector, and the stau mixing angle $\theta_{\tilde\tau}$,
we expand  the imaginary part of the product
of $\stau_m$-$\tau$-$\tilde\chi_i^0$ couplings
\begin{eqnarray}
  \Im \{ a^{\stau}_{mi} (b^{\stau}_{mi})^\ast \}  
&=&  
\Im \Big\{|\mathcal{R}_{m1}|^2 f^{L     }_{\tau i} h^{R}_{\tau i} +
      |\mathcal{R}_{m2}|^2 f^{\ast R}_{\tau i} h^{R}_{\tau i} \nonumber\\ 
 & +& \mathcal{R}_{m1}\mathcal{R}_{m2}^\ast 
      \big[(h^{R}_{\tau i})^2 - f^{R}_{\tau i} f^{\ast L}_{\tau i}\big]
 \Big\},
\end{eqnarray}
in terms of the stau mixing matrix $\mathcal{R}$, 
the  gauge couplings $f_{\tau i}^{L,R}$ and the higgs 
couplings $h_{\tau i}^{L,R}$.
In particular, for a CP-conserving neutralino sector,
$\phi_1=\phi_\mu=0$, we have
\begin{eqnarray}
  \Im \{ a^{\stau}_{mi} (b^{\stau}_{mi})^\ast \}  =
 \,^{\;\,+}_{(-)}
  \sin\phi_{\tilde\tau} \sin(2 \theta_{\tilde\tau})
     \frac{1}{2} 
  \left[(h^{R}_{\tau i})^2 - f^{R}_{\tau i} f^{L}_{\tau i}\right],
\nonumber\\
\label{eq:expectedeta}
\end{eqnarray}
for $m=1$, and the sign in parentheses holds for $m=2$.
Thus we expect a maximal $\eta$ and thus maximal asymmetries
for maximal stau 
mixing\footnote{
  Note that a maximal mixing is naturally achieved for nearly degenerate staus. 
  However then the asymmetries for $\tilde\tau_1$ and $\tilde\tau_2$ 
  decay typically have similar magnitude but opposite sign, 
  and thus might cancel. See the discussion at the end of the numerics
  in Section~\ref{sub:MLMR}.
},
$\theta_{\tilde\tau}\approx \pm\pi/4$,
and a maximal CP phase in the stau mixing matrix,
$\phi_{\tilde\tau}\approx \pm\pi/2$.
Note that, in particular, the dependence of  $\phi_{\tilde\tau}$
on  $\phi_{A_\tau}$ is strong for
$|A_\tau| > |\mu| \tan\beta$.
We will study numerically the phase and parameter dependence on 
$\ACP$ and $\eta$ further in Section~\ref{sec:results}.

\subsection{Boost dependence}\label{sub:boost}
\begin{figure}[t!]  
  \label{fig:LHCboosts}
  \includegraphics[angle =0,width=0.74\columnwidth]{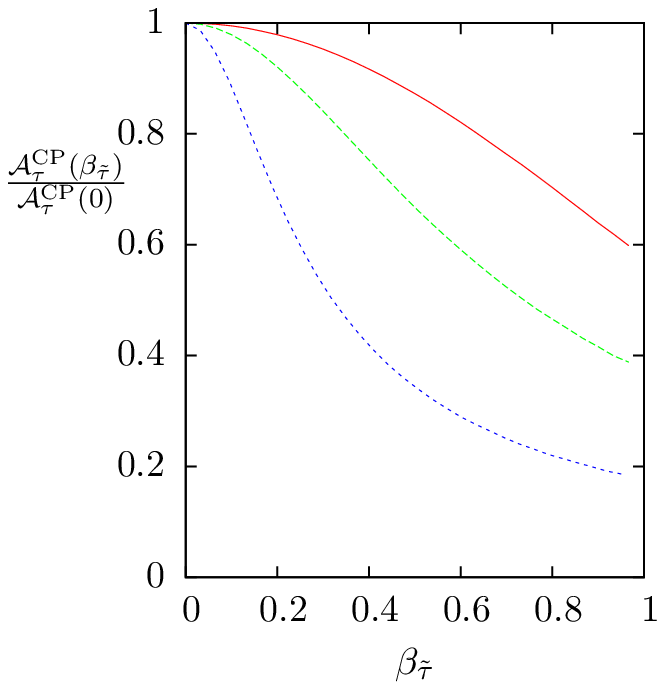}
  \caption{Boost distributions of the $\tau$ polarization asymmetry
          $\ACP$, Eq.~(\ref{eq:asym}),
          normalized by $\ACP(\beta_{\tilde\tau}=0)$,
          for three different sets of stau masses,
          $m_{\tilde\tau_{1,2}} \approx 200$~GeV (solid, red),
          $400$~GeV (dashed, green), and
          $1000$~TeV (dotted, blue), see text, for
          stau decay  $\tilde\tau_1   \to \tau  \tilde\chi_2^0$,
           followed by
          $\tilde\chi_2^0 \to \ell_1 \tilde\ell_R$, and 
          $\tilde\ell_R   \to \tilde\chi^0_1 \ell_2$
          ($\ell= e$ or $\mu$), see Fig.~\ref{Feyn},
          The SUSY parameters are given in 
          Table~\ref{tab:benchmark.scenario}.
  }
\end{figure}

The triple product asymmetry 
$\ACP$, Eq.~(\ref{P2}),
is not Lorentz invariant but depends on the boost of the decaying stau,
\begin{equation}
\beta_{\tilde\tau}=\dfrac{|{\bf p}_{\tilde\tau}|}{E_{\tilde\tau}}.
\end{equation}
In Fig.~2, we show the boost dependence
of the asymmetry $\ACP(\beta_{\tilde\tau})$,
normalized by $\ACP(\beta_{\tilde\tau}=0)$.
The SUSY parameters are given in Table~\ref{tab:benchmark.scenario},
and we have chosen three sets  of  different $\tilde\tau$ 
soft-breaking parameters
$\{M_{{\tilde{E}}_\tau}, M_{{\tilde{L}}_\tau} \}=
 \{ 195,200 \}$~GeV (solid, red);
$\{ 395,400 \}$~GeV (dashed, green); and
$\{ 998,1000 \}$~GeV (dotted, blue).
The corresponding stau masses are
$\{m_{\tilde\tau_1}, m_{\tilde\tau_2} \} =
\{ 194,209 \}$; $\{ 395,404 \}$; $\{ 998,1002 \}$~GeV, respectively.
The corresponding asymmetries in the stau rest frame are
$\ACP(\beta_{\tilde\tau})= -66\%$; $-72\%$, $-71\%$.
Note that we have chosen nearly degenerate stau masses which lead to an 
enhanced stau mixing and thus to maximal asymmetries; 
see also the discussion in Section~\ref{sec:results}.

\medskip

For the stau masses 
$\{m_{\tilde\tau_1}, m_{\tilde\tau_2} \} =\{ 194,209 \}$~GeV,
the staus can be produced at the ILC with $\sqrt{s}=500$~GeV, and have
a fixed boost of $\beta_{\tilde\tau}= 0.63$. The corresponding
asymmetry is then reduced to $\ACP=-53\%$
if the stau rest frame cannot be reconstructed.
Typical ILC cross section for these masses 
are of the order of some $20$~fb~\cite{Alwall:2007st}.
\medskip

If the staus are produced at the LHC, they will have a distinct boost 
distribution depending on their mass,  which typically peaks at 
high values $\beta_{\tilde\tau}\approx 0.9$ for stau masses of 
the order of a few $100$~GeV up to a $1$~TeV, see e.g. 
Refs~\cite{Deppisch:2009nj,Deppisch:2010nc}.
Then the normal tau polarization in the laboratory frame is  
obtained by folding the boost dependent polarization 
$\ACP$ with the normalized stau boost 
distribution~\cite{Deppisch:2009nj},
\begin{equation}
 \ACP_{\rm lab}=
   \dfrac{1}{\sigma_P}\int_0^1\dfrac{{\rm d}\sigma_P}
  {{\rm d}\beta_{\tilde\tau}}\ACP(\beta_{\tilde\tau})\, 
   {\rm d}\beta_{\tilde\tau}, 
\end{equation}
with the production cross section 
$\sigma_P=\sigma(pp\to\tilde\tau^+\tilde\tau^-)$. 
The typical reduction of the normal tau
polarization $\ACP_{\rm lab}$ is of the order of two thirds of
the asymmetry compared to that in the stau rest frame $\ACP(0)$.
However, it has been recently shown (for similar asymmetries in
stop decays at the LHC), that the rest frame can be partly reconstructed
event by event using on-shell mass conditions,
see Refs.~\cite{MoortgatPick:2009jy}.
The  LHC cross section for stau pair production,
$ \sigma(pp\to\tilde\tau_1^+\tilde\tau_1^-)$,
also sensitively depends on the stau masses, e.g.,
for our benchmark scenario in Table~\ref{tab:benchmark.scenario}, 
we find cross sections up to $10$~fb at $\sqrt{s}=14$~TeV~\cite{Alwall:2007st}.

\begin{table}[t!]  
\caption{
Benchmark scenario.
The mass parameters $M_2$, $|\mu|$, $A_\tau$, $M_{\tilde E}$, $M_{\tilde L}$  
        $M_{{\tilde E}_\tau}$, and $M_{{\tilde L}_\tau}$ are given in GeV.
}  
\begin{tabular}{cp{0.02\columnwidth}cp{0.02\columnwidth}cp{0.03\columnwidth}
                cp{0.03\columnwidth}cp{0.03\columnwidth}cp{0.02\columnwidth}
                c}
\\  
\hline\hline
${{\phi_{1}}_{\phantom{I}}}_{\phantom{I}}$&&
${{\phi_{\mu}}_{\phantom{I}}}_{\phantom{I}}$&&
${{\phi_{A_\tau}}_{\phantom{I}}}_{\phantom{I}}$&&
${M_2}_{\phantom{I}}$&&
${|\mu|}_{\phantom{I}}$&&
${A_{\tau}}_{\phantom{I}}$&&
${\tan\beta}_{\phantom{I}}$ 
\\ 
\hline  
0&&
0&&
$\pi/2$&&
250&&
250&&
2000&&
3
\\  
\hline\hline 
${M_{{\tilde E}_\tau}}_{\phantom{I}}$&&&&
${M_{{\tilde L}_\tau}}_{\phantom{I}}$&&&&
${M_{\tilde E}}_{\phantom{I}}$&&&&
${M_{\tilde L}}_{\phantom{I}}$\\
\hline
495&&&&
500&&&&
150&&&&
200\\
\hline\hline 
\end{tabular}  
\label{tab:benchmark.scenario}  
\end{table}

\begin{table}[b!]  
\caption{Mass spectrum for the scenario in 
        Table~\ref{tab:benchmark.scenario}.}
\begin{tabular}{cp{0.03\columnwidth}cp{0.05\columnwidth}cp{0.09\columnwidth}  
                cp{0.06\columnwidth}cp{0.06\columnwidth}cp{0.06\columnwidth}c}
\hline\hline  
${\tilde\ell}$             && $m$~[GeV]&&&&&&&&$\tilde\chi$&&$m$~[GeV]\\  
\hline    
${\tilde e}_R,{\tilde \mu}_R$  && 155     &&&&&&&&${\tilde\chi}_1^0$&& 112\\  
${\tilde e}_L,{\tilde \mu}_L$  && 204     &&&&&&&&${\tilde\chi}_2^0$&& 190\\  
${\tilde\nu}_e,{\tilde\nu}_\mu$&& 192    &&&&&&&&${\tilde\chi}_3^0$&& 254\\  
${\tilde\nu}_\tau$           && 497    &&&&&&&&${\tilde\chi}_4^0$&& 327\\
${\tilde\tau}_1$             && 495   &&&&&&&&${\tilde\chi}_1^\pm$&& 181\\ 
${\tilde\tau}_2$             && 504   &&&&&&&&${\tilde\chi}_2^\pm$&& 325\\
\hline\hline  
\end{tabular}  
\label{tab:benchmark.masses}  
\end{table}

\section{Numerical results}\label{sec:results}

We quantitatively study  the tau polarization 
asymmetry, and the  branching ratios for the two-body decay chain 
\begin{equation}
 \tilde\tau_1   \to \tau + \tilde\chi_2^0;  \quad 
 \tilde\chi_2^0 \to \ell_1^+ + \tilde\ell_R^-;  \quad 
 \tilde\ell_R^-   \to \tilde\chi^0_1 + \ell_2^-,
\end{equation}
for $\ell= e,\mu$.
The asymmetry  probes the  MSSM phases 
$\phi_1$, $\phi_\mu$ and $\phi_{A_\tau}$,
of the neutralino and stau sector. 
We center our numerical discussion around a general MSSM benchmark 
scenario, see Table~\ref{tab:benchmark.scenario}.
We choose heavier soft breaking parameters 
in the stau sector than in the $\tilde e,\tilde \mu$ sector,
to enable the mass hierarchy
\begin{equation}
m_{\tilde\tau_m} > m_{ \tilde\chi_i^0} > m_{\tilde\ell_R} > m_{
  \tilde\chi_1^0}. 
\end{equation}
Further we choose almost degenerate staus which enhances
their mixing, leading to maximal asymmetries.
We choose a large value of the trilinear scalar
coupling parameter,  $|A_\tau| > |\mu| \tan\beta$\footnote{The value of $|A_\tau|$ is restricted bz the vacuum stability contidtion as $|A_\tau^2< 3(m_{\tilde\tau}^2+m_{{\tilde\nu}_\tau}^2+M_H^2+\mu^2|$~\cite{Frere:1983}.},
to enhance the impact of $\phi_{A_\tau}$ in the stau sector.
Finally, to reduce the number of MSSM parameters, we use the (GUT inspired) 
relation $|M_1|=5/3\,M_2\tan^2 \theta_w$~\cite{Haber:1984rc}
for the gaugino mass parameters. 
The resulting masses of the staus, neutralinos and charginos  are
summarized in Table~\ref{tab:benchmark.masses}.

\subsection{Phase dependence}\label{sub:phases}
\begin{figure*}[t!]  
\subfigure[]{  
  \includegraphics[width=0.96
\columnwidth]{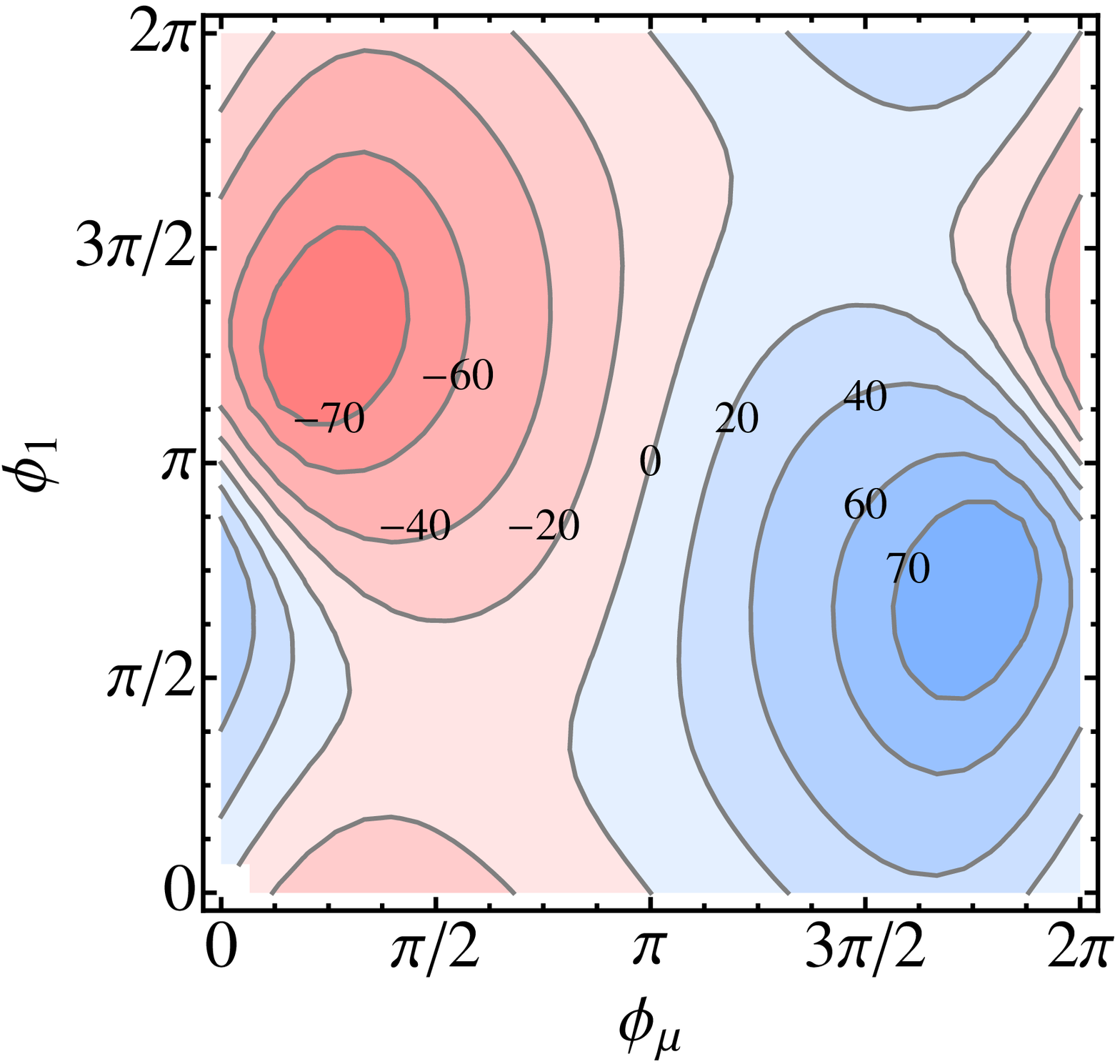}
 \label{fig:phimu_m1_asym}}
  \hfill  
\subfigure[]{
  \includegraphics[width=0.96
\columnwidth]{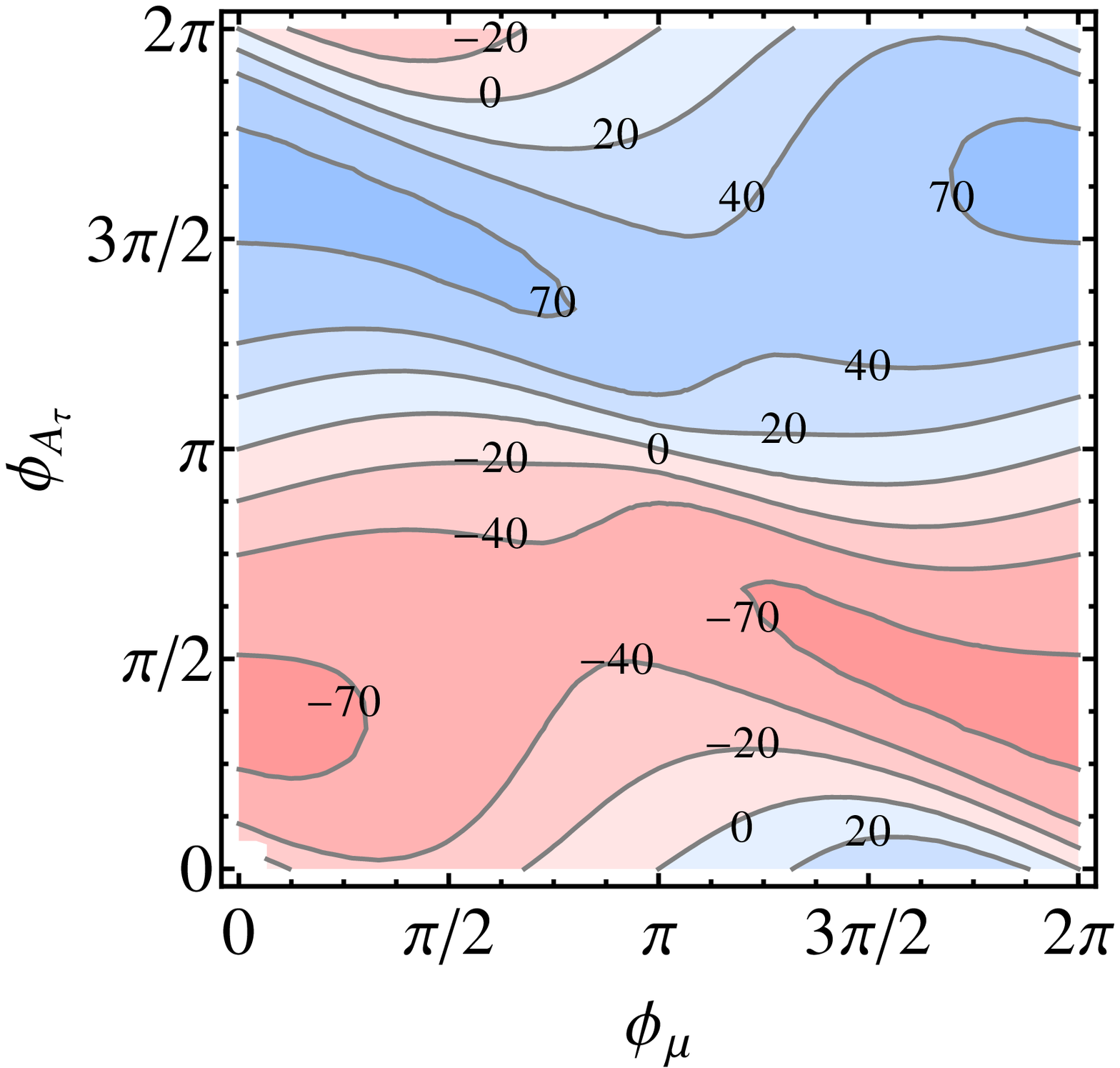}
  \label{fig:phimu_Atau_asym}}
  \caption{\label{fig:phimu_Atau_m1}
      Phase dependence of (a) the $\tau$ polarization asymmetry 
      $\ACP$, Eq.~(\ref{eq:asym}), in percent,
      in the $\phi_1$--$\phi_{\mu}$ plane (for $\phi_{A_\tau}=0$),
      and (b) in the $\phi_{A_\tau}$--$\phi_\mu$ plane
       (for $\phi_{1}=0$),
      in the stau rest frame. We consider the decay
      $\tilde\tau_1 \to \tau  \tilde\chi_2^0$,
      followed by
      $\tilde\chi_2^0 \to \ell_1^+ \tilde\ell_R^-$, and 
      $\tilde\ell_R^-   \to \tilde\chi^0_1 \ell_2^-$
      where $\ell= e$ or $\mu$, {\it cf.} Fig.~\ref{Feyn}.
     The other MSSM parameters are defined in
     Table~\ref{tab:benchmark.scenario}.
   } 
\end{figure*}

For the benchmark scenario given in Table~\ref{tab:benchmark.scenario}, 
we study the phase dependence of the asymmetry $\ACP$ in the stau 
rest frame.
In Fig.~\ref{fig:phimu_m1_asym},  we show the dependence on the
CP phases in the neutralino sector, $\phi_{1}$ and $\phi_\mu$.
In Fig.~\ref{fig:phimu_Atau_asym}, we show the dependence on
the phases in the stau secotor  $\phi_{A_\tau}$ and $\phi_\mu$.
The asymmetry strongly depends on 
$\phi_{A_\tau}\approx\phi_{\tilde\tau}$, which
we expect for $|A_\tau| \gg |\mu| \tan\beta$ as in our
benchmark scenario, see Table~\ref{tab:benchmark.scenario}.
In particular for $\phi_\mu=0$ in Fig.~\ref{fig:phimu_Atau_asym},
the asymmetry follows the approximation formula Eq.~\eqref{eq:expectedeta}, 
and attains its maximal values at
$\phi_{\tilde{\tau}}\approx\phi_{A_\tau}\approx \pm \pi/2$.

\subsection{{\boldmath $|A_\tau|$}--{\boldmath $\tan\beta$} dependence
  and       stau mixing}
                                                  \label{sub:AtauTanb}
\begin{figure*}[ht!]  
\subfigure[]{  
 \includegraphics[width=0.97\columnwidth]{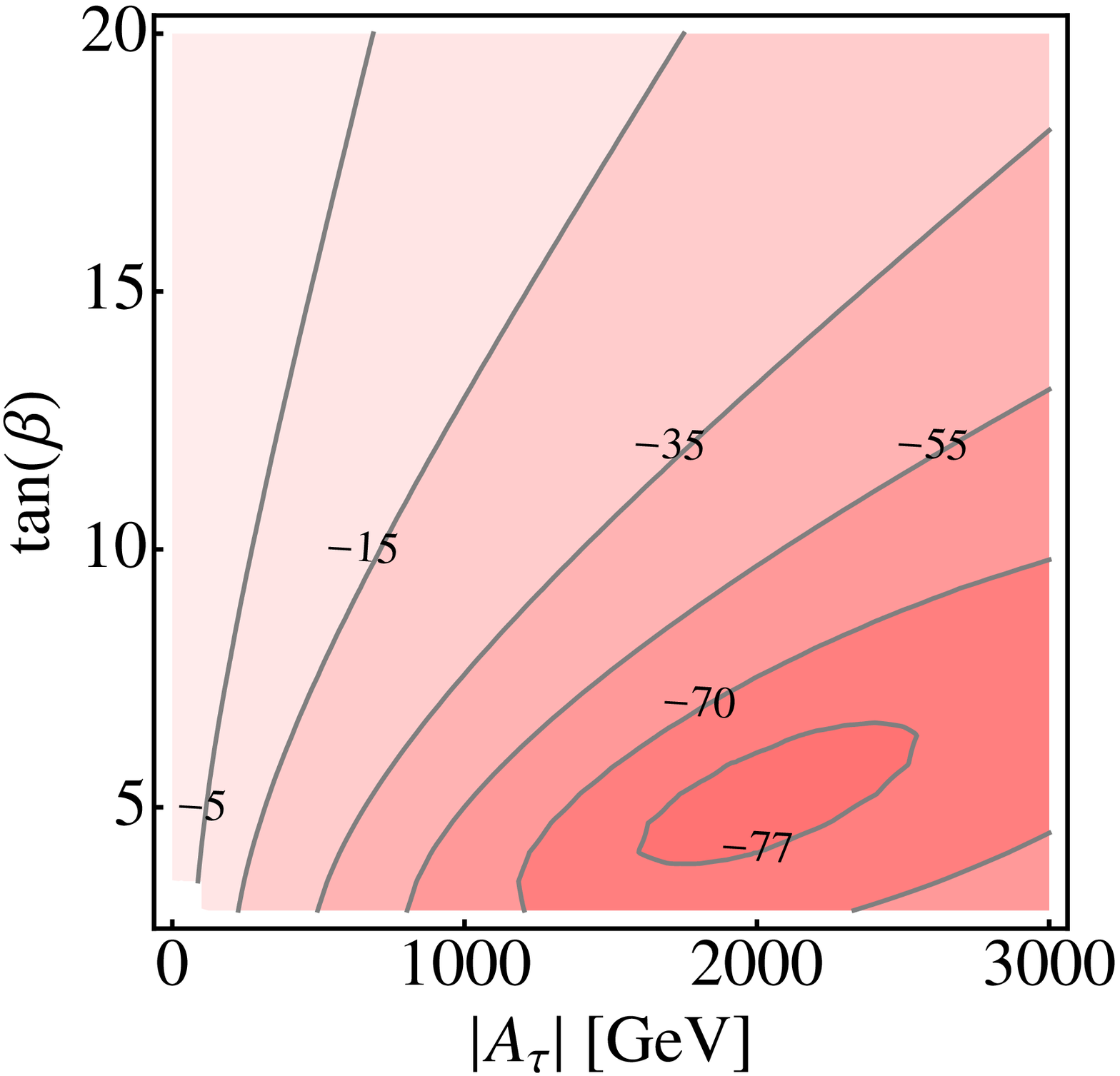}
  \label{fig:A_tanbeta_asym}}
  \hfill  
\subfigure[]{  
  \includegraphics[width=0.97\columnwidth]{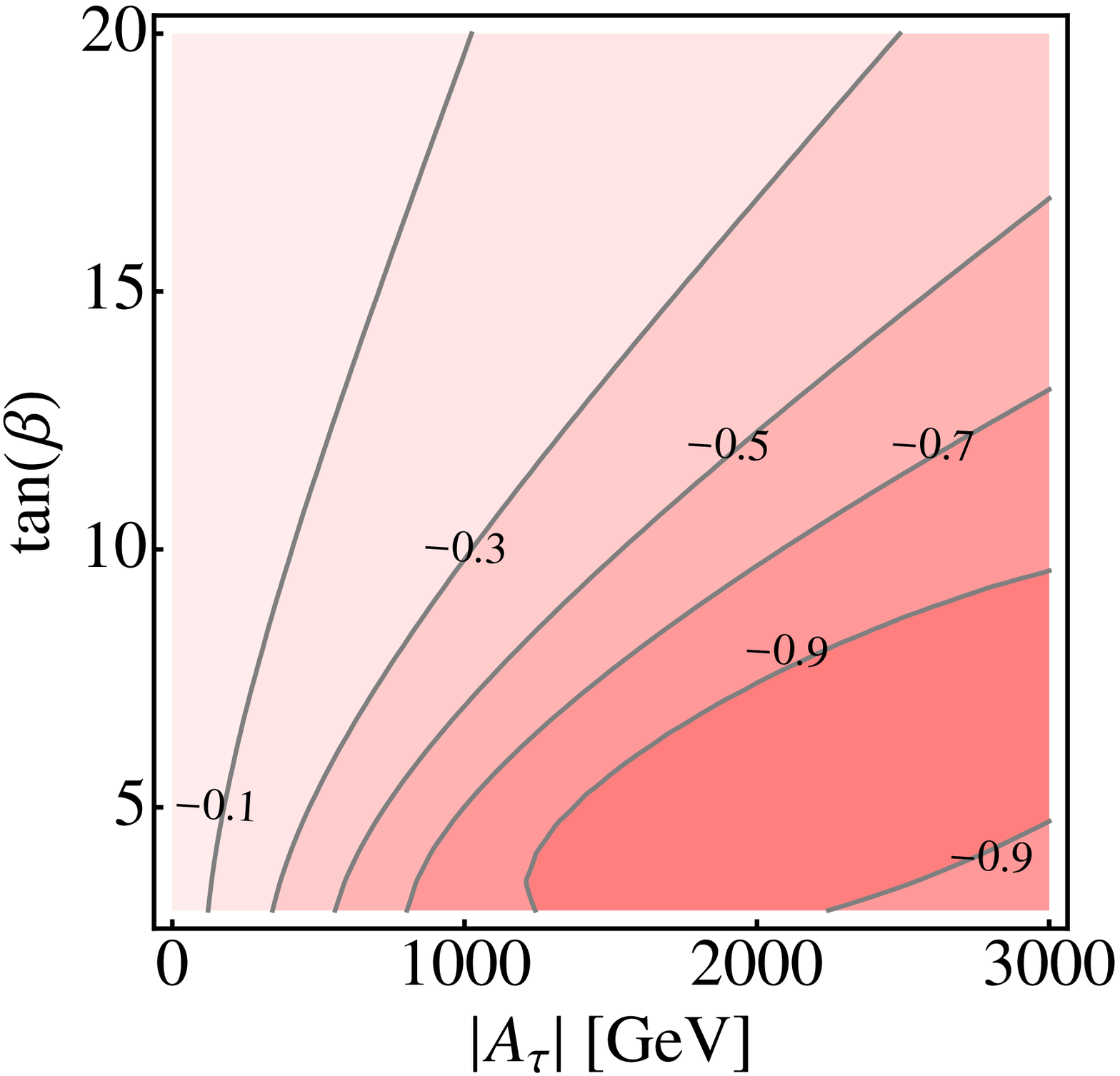}
\label{fig:A_tanbeta_eta_pi4}}
\subfigure[]{
  \includegraphics[width=0.97\columnwidth]{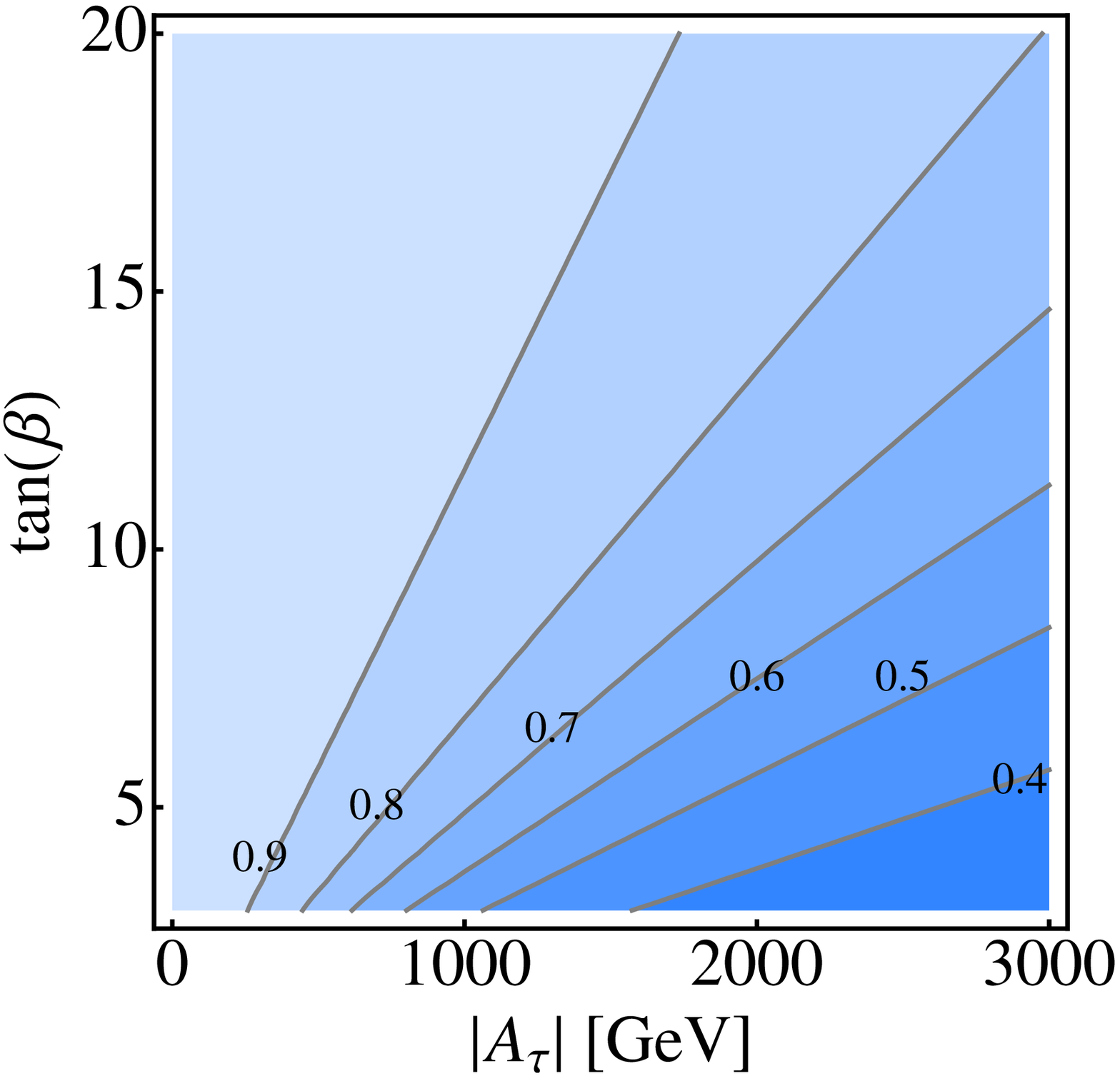}
  \label{fig:A_tanbeta_phistau_pi4}}
  \hfill  
\subfigure[]{  
\includegraphics[width=0.97\columnwidth]{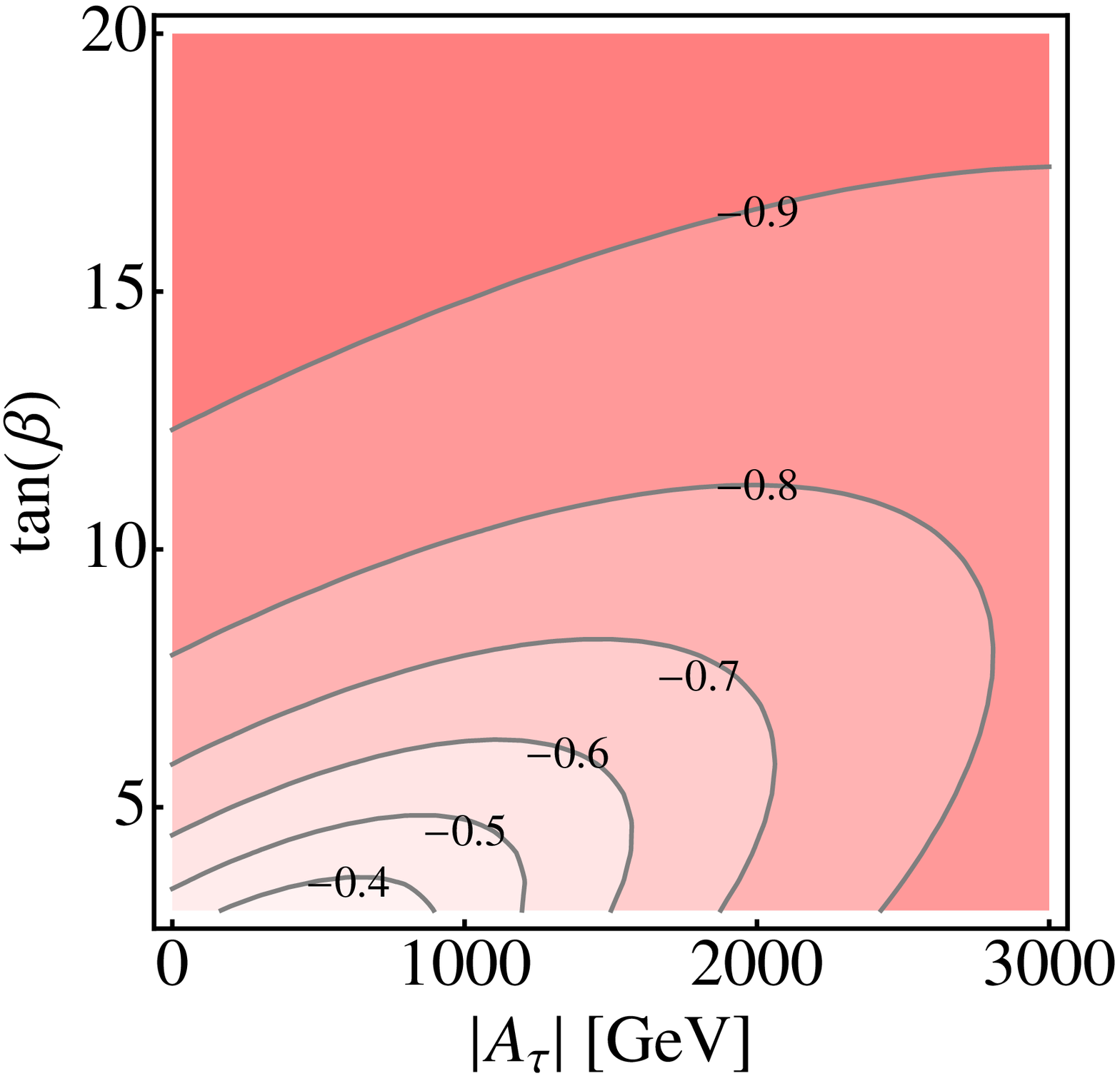}
  \label{fig:A_tanbeta_cosstau_pi4}}
  \caption{ \label{fig:A_tanbeta} 
      $|A_{\tau}|$--$\tan\beta$ dependence of
      (a)~the $\tau$ polarization asymmetry 
      $\ACP$, Eq.~(\ref{eq:asym}), in percent,
      in the stau rest frame (for the decay
      $\tilde\tau_1 \to \tau  \tilde\chi_2^0$,
      followed by
      $\tilde\chi_2^0 \to \ell_1^+ \tilde\ell_R^-$, and 
      $\tilde\ell_R^-   \to \tilde\chi^0_1 \ell_2^-$
      for $\ell= e$ or $\mu$, {\it cf.} Fig.~\ref{Feyn}),
      (b)~the coupling factor $\eta$, Eq.~(\ref{etafactor}),
      (c)~the phase  $\phi_{\tilde\tau}$ in the stau sector,
         Eq.~(\ref{eq:phi_stau}), and
      (d)~$\sin(2\theta_{\tilde\tau})$, with $\theta_{\tilde\tau}$
          the stau mixing angle,  
          Eqs.~(\ref{eq:cos_theta_stau}), (\ref{eq:sin_theta_stau}).
         The plots are for $\phi_{A_\tau} = \pi/4,$ the other 
          MSSM parameters are given in 
         Table~\ref{tab:benchmark.scenario}.
}
\end{figure*}  
In Fig.~\ref{fig:A_tanbeta_asym}, we show the $|A_\tau|$ and $\tan\beta$ 
dependence of the asymmetry $\ACP$ in the stau rest frame.
We can observe that the asymmetry obtains its maximum, 
$\ACP\approx-77\%$, where also the coupling factor is maximal,
 $\eta \approx 0.95$, see Fig.~\ref{fig:A_tanbeta_eta_pi4}.  
As discussed in Subsection~\ref{sub:analytics}, 
the imaginary part of the product of the stau couplings
$\Im \{ a^{\stau}_{mi} (b^{\stau}_{mi})^\ast \}$ 
is maximal for a maximal CP phase  $\phi_{\tilde\tau}=\pi/2$ in
the stau sector, which we show in Fig.~\ref{fig:A_tanbeta_phistau_pi4}.
Note that the location of the  maximum of $\ACP$
is not at maximal stau mixing,
$\sin(\theta_{\tilde\tau}) = 1/\sqrt{2}\approx 0.7$, since 
$\eta\propto\sin(2\theta_{\tilde\tau})/(|a^{\tilde\tau}|^2 +|b^{\tilde\tau}|^2)$ 
starts to decrease for increasing  $A_\tau$ and $\tan\beta$. 

\smallskip

To study the stau mixing,
we show the $M_{{\tilde{E}}_\tau}$--$M_{{\tilde{L}}_\tau}$
dependence of the asymmetry
$\ACP$ in  Fig.~\ref{fig:ML_MR_asym}.
In the  entire $M_{{\tilde{E}}_\tau}$--$M_{{\tilde{L}}_\tau}$
plane, the CP phase in the stau sector is almost
maximal, $\phi_{\tilde\tau}=0.61 \pi$.
However, the asymmetry  obtains its maxima in
the small corridor $M_{{\tilde{E}}_\tau} \approx M_{{\tilde{L}}_\tau}$,
where the stau mixing is maximal, 
$\theta_{\tilde\tau} = \pi/4$.\vspace*{-2ex}
\subsection{{\boldmath $|\mu|$}--{\boldmath $M_2$} dependence and
  branching   ratios}  
                                                       \label{sec:muM2}
We show the $|\mu|$--$M_2$ dependence of the asymmetry
$\ACP$ in  Fig.~\ref{fig:mu_m2_asym}.
The maxima of $\ACP$ are obtained where  
the coupling factor $\eta$
is also maximal, see Eq.~(\ref{etafactor}).

\smallskip

In Fig.~\ref{fig:mu_m2_BR_stau}, we show the corresponding 
stau branching ratio,
${\rm BR}(\tilde\tau_1\to\tau\tilde\chi_2^0)$, 
which can be as large as $40$\%. Other competing channels 
can reach
${\rm BR}(\tilde\tau_1\to\tau\tilde\chi_{1}^0)\approx 65\%$,
and 
${\rm BR}(\tilde\tau_1\to\nu_\tau\tilde\chi_{1(2)}^\pm)\approx 20(10)\%$.
The stau decay into the chargino $\tilde\chi_1^\pm$ is always open 
since typically the second lightest neutralino and the lightest chargino are
almost degenerate, $m_{\tilde\chi_2^0} \approx  m_{\tilde\chi_1^\pm}$.
The neutralino branching ratio
${\rm BR}(\tilde\chi_2^0 \to \ell \tilde\ell_R)$,
summed over $\ell=e,\mu$, is shown in Fig.~\ref{fig:mu_m2_BR_chi},
which reaches up to $100$\%. The other important competing 
decay channels are 
${\rm BR}(\tilde\chi_2^0 \to \nu_\ell \tilde\nu_\ell)$,
and
${\rm BR}(\tilde\chi_2^0 \to \ell \tilde\ell_L)$,
which open around $\mu\approx 250$~GeV and  $\mu\approx 300$~GeV, respectively, 
for  $M_2 = 250$~GeV.
Note that in our benchmark scenario, see 
Table~\ref{tab:benchmark.scenario}, we have
${\rm BR}(\tilde\ell_R \to \tilde\chi^0_1 \ell)=1$.


\subsection{Impact of {\boldmath $\tilde{\tau}_2$} decay }
                                         \label{sub:MLMR}
As we discussed in Section~\ref{sub:AtauTanb}, we find large asymmetries for 
nearly degenerate staus, where we naturally obtain a maximal stau mixing.
However, then typically the asymmetries for $\tilde\tau_1$ and $\tilde\tau_2$ 
decay are similar in magnitude, but opposite in sign. For example in our 
benchmark scenario we find $\ACP=-71\%$ for $\tilde\tau_1$ decay, but  
$\ACP=+32\%$ for the decay of $\tilde\tau_2$. If the production and decay 
process of $\tilde\tau_1$ 
cannot be experimentally disentangled from that of $\tilde\tau_2$ properly, 
the two asymmetries might cancel. We show their sum 
in Fig.~\ref{fig:ML-MR_asymm_k1-plus-k2} in the 
$M_{{\tilde{E}}_\tau}$--$M_{{\tilde{L}}_\tau}$ plane.
In Fig.~\ref{fig:ML-MR_Mstau2-Mstau1}, we show the corresponding
stau mass splitting.

\smallskip

\begin{figure*}[h!]  
\subfigure[]{  
\includegraphics[width=0.97\columnwidth]{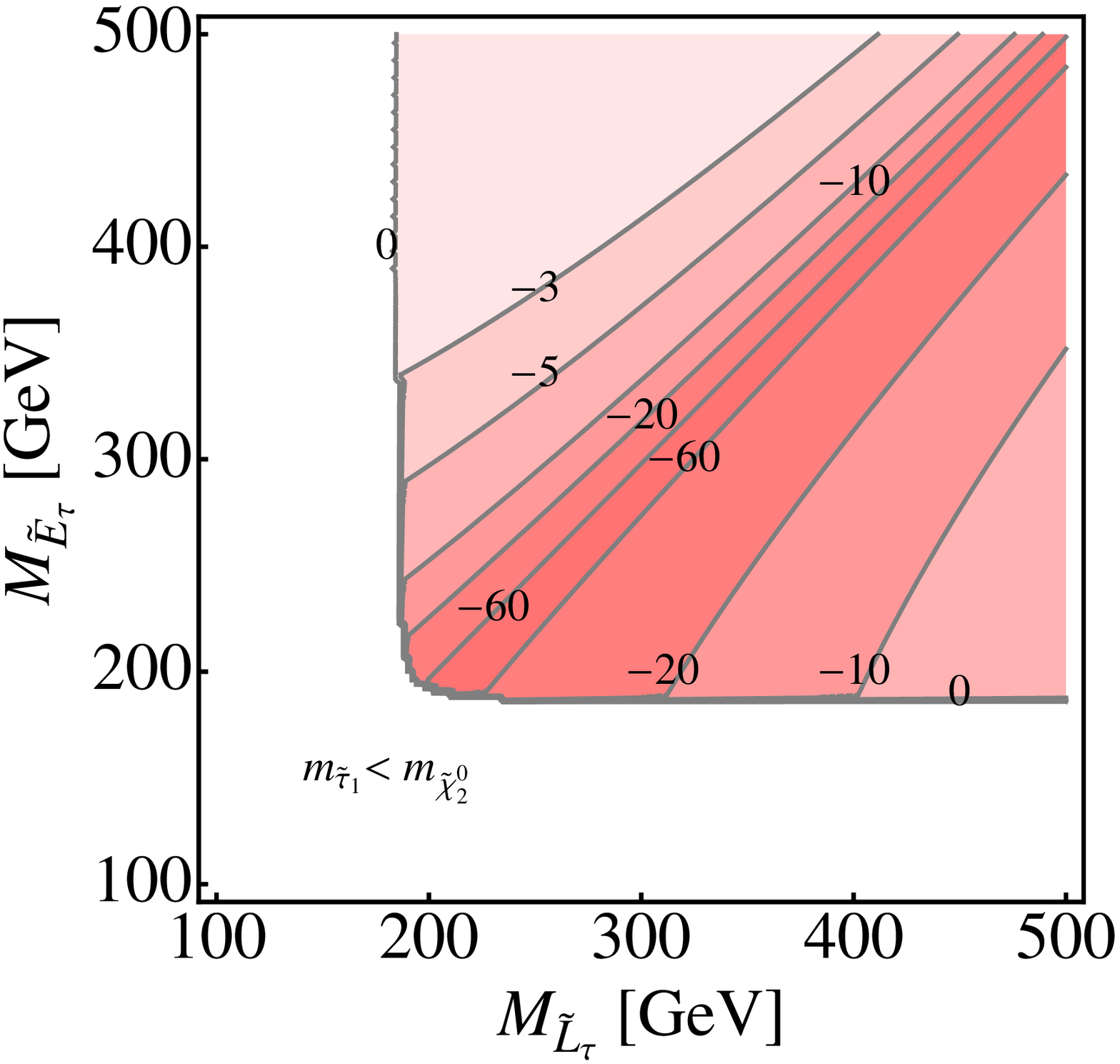}
\label{fig:ML_MR_asym}}
\subfigure[]{  
  \includegraphics[width=0.97\columnwidth]{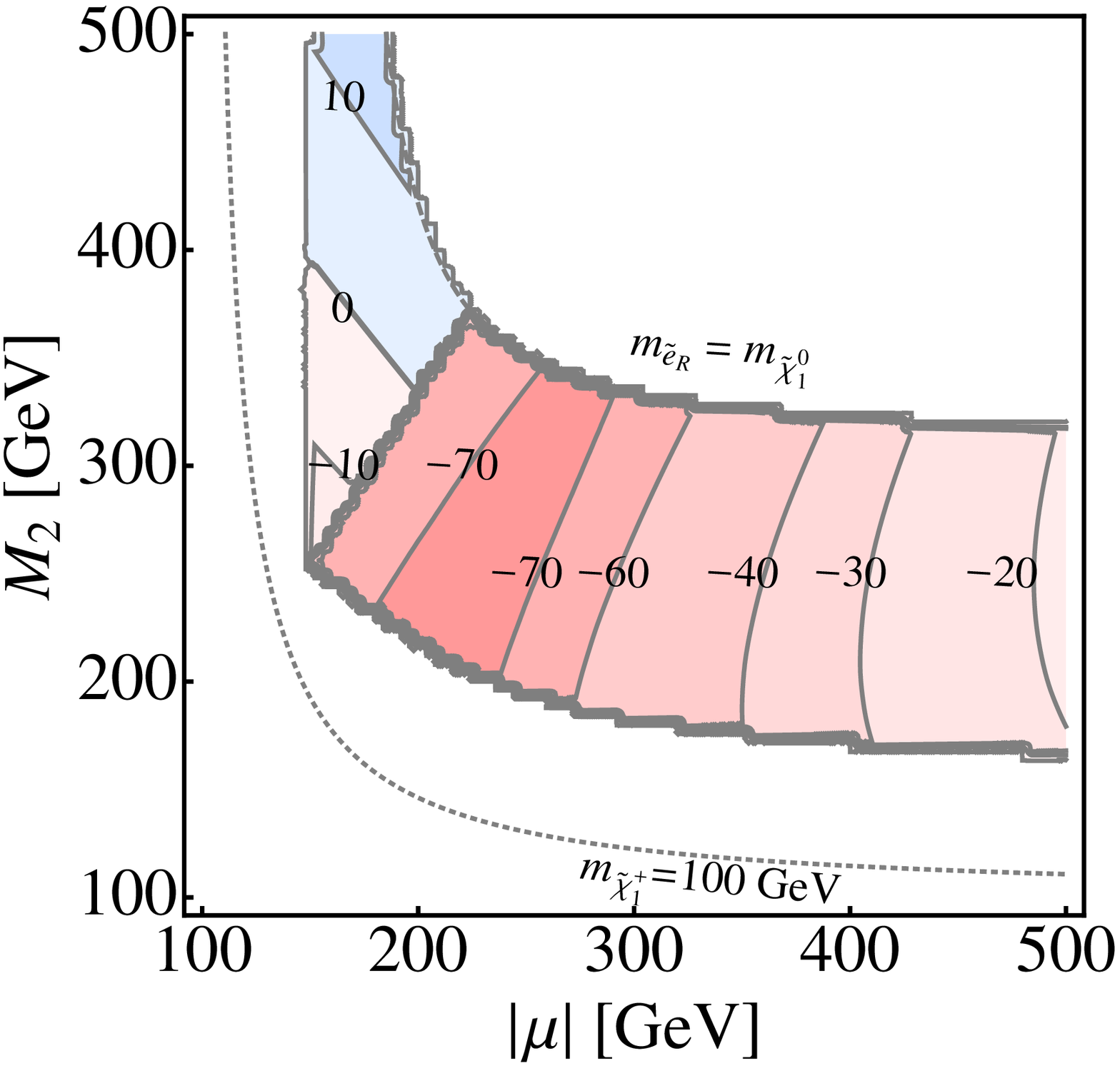}
  \label{fig:mu_m2_asym}}
  \hfill  
 \caption{\label{fig:ML_MR_}
     Dependence of the $\tau$ polarization asymmetry 
      $\ACP$, Eq.~(\ref{eq:asym}), in percent,
      in the stau rest frame (for the decay
      $\tilde\tau_1 \to \tau  \tilde\chi_2^0$,
      followed by
      $\tilde\chi_2^0 \to \ell_1^+ \tilde\ell_R^-$, and 
      $\tilde\ell_R^- \to \tilde\chi^0_1 \ell_2^-$
      for $\ell= e$ or $\mu$, see Fig.~\ref{Feyn}), on
      (a)~the soft breaking parameters in the stau sector
      $M_{{\tilde E}_\tau}$,  $M_{{\tilde L}_\tau}$,  Eqs.~(\ref{eq:left_stau}),  
                                      Eqs.~(\ref{eq:right_stau}).
       In (b)~the dependence of $\ACP$ on the
      gaugino and higgsino parameters  $|\mu|$, $M_2$.
      Below the contour $m_{\tilde e_R} = m_{\tilde\chi_2^0}$
      the two-body decay $\tilde\chi_2^0 \to \ell \tilde\ell_R$ 
      is kinematically forbidden,
      above the contour $m_{\tilde  e_R} = m_{\tilde\chi_1^0}$
      the lightest neutralino is no longer the LSP since 
      $m_{\tilde e_R}<m_{\tilde\chi_1^0}$.
       Below the contour $ m_{\tilde\chi_1^\pm}=100$~GeV the lightest
      chargino is lighter than $100$~GeV.
    The MSSM parameters are given in Table~\ref{tab:benchmark.scenario}.
} 
\end{figure*}

\begin{figure*}[h!]  
\subfigure[]{ \label{fig:mu_m2_BR_stau}
\includegraphics[width=0.97\columnwidth]{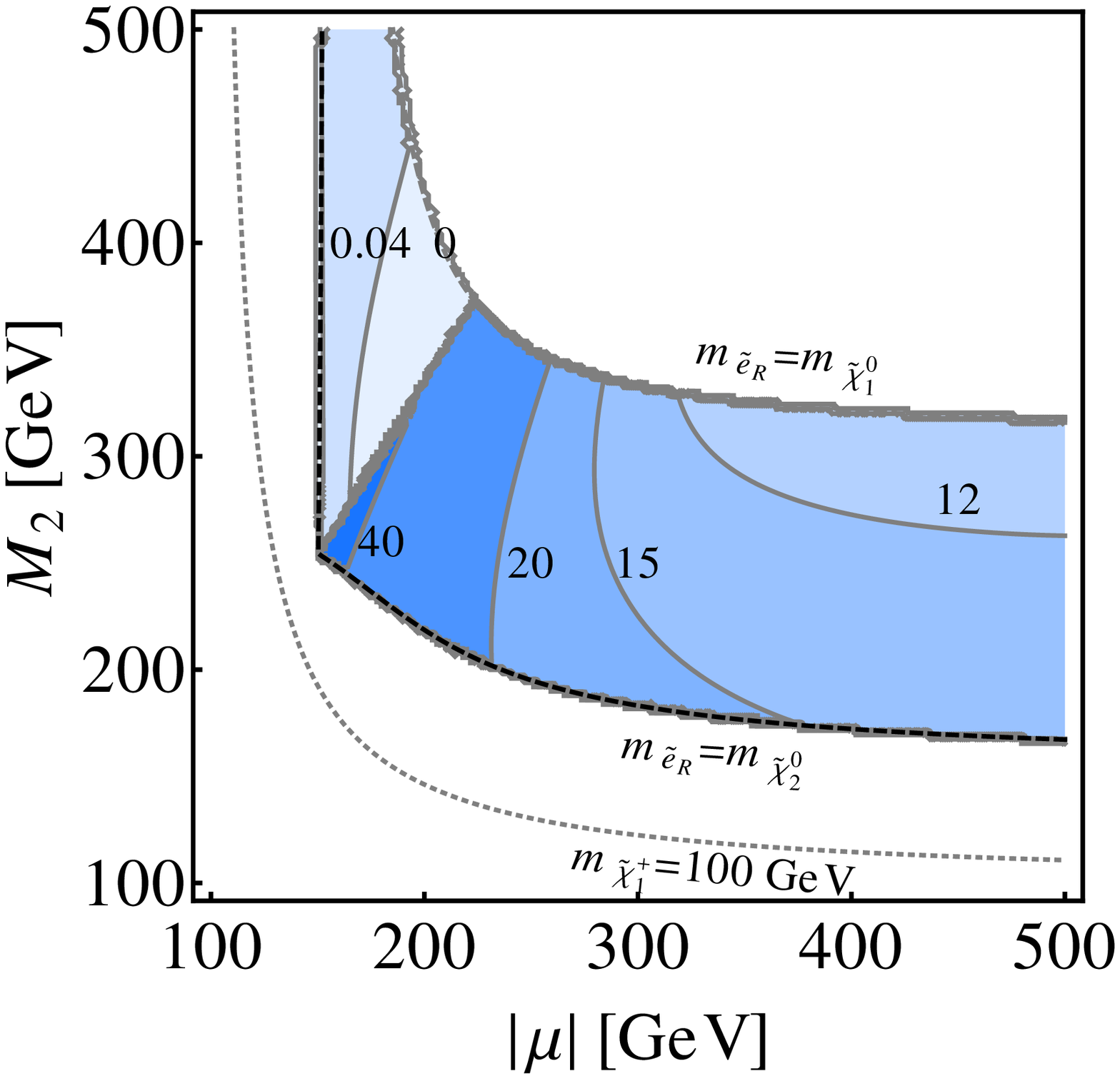}}    
   \hfill  
\subfigure[]{\label{fig:mu_m2_BR_chi}
\includegraphics[width=0.97\columnwidth]{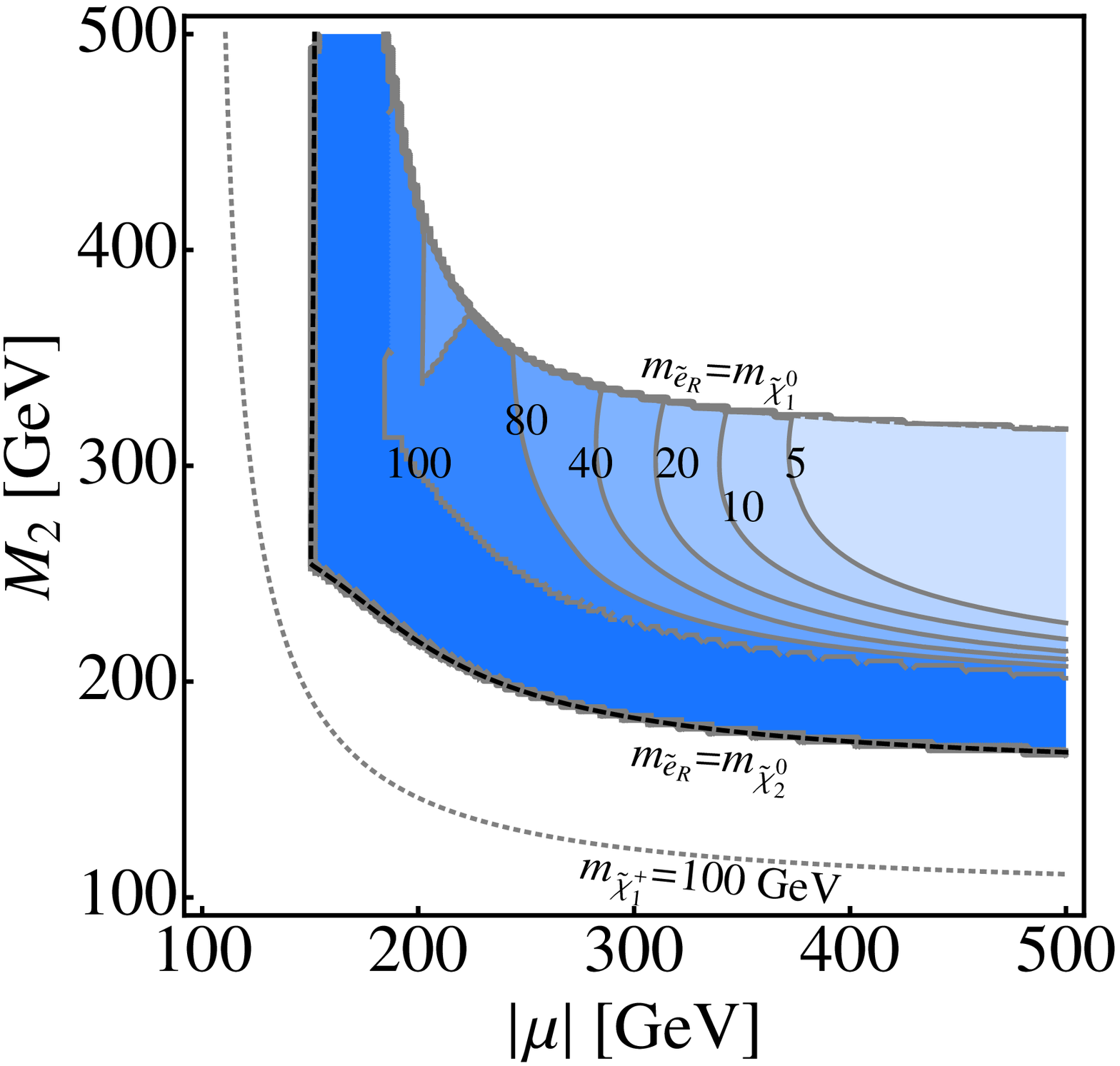}}
  \caption{
     Contour lines in the $|\mu|$--$M_2$ plane of
     (a)~the stau branching ratio
     ${\rm BR}(\tilde\tau_1\to\tau\tilde\chi_2^0)$ in percent, and
     (b)~the neutralino branching ratio
     ${\rm BR}(\tilde\chi_2^0 \to \ell \tilde\ell_R)$, in percent,
      summed over both lepton flavors $\ell =e$, $\mu$ and charges, for the 
      MSSM parameters as given in Table~\ref{tab:benchmark.scenario}.
      Below the contours $m_{\tilde e_R} = m_{\tilde\chi_2^0}$
      in Figs.~\ref{fig:mu_m2_BR_stau}, \ref{fig:mu_m2_BR_chi},
      the two-body decay $\tilde\chi_2^0 \to \ell \tilde\ell_R$ 
      is kinematically forbidden,
      above the contours $m_{\tilde e_R} = m_{\tilde\chi_1^0}$
      the lightest neutralino is no longer the LSP since 
      $m_{\tilde e_R}<m_{\tilde\chi_1^0}$.
       Below the contours $ m_{\tilde\chi_1^\pm}=100$~GeV the lightest
      chargino is lighter than $100$~GeV.
     } 
\end{figure*}

\begin{figure}[t!]  
\subfigure[]{  
  \includegraphics[width=0.95\columnwidth]{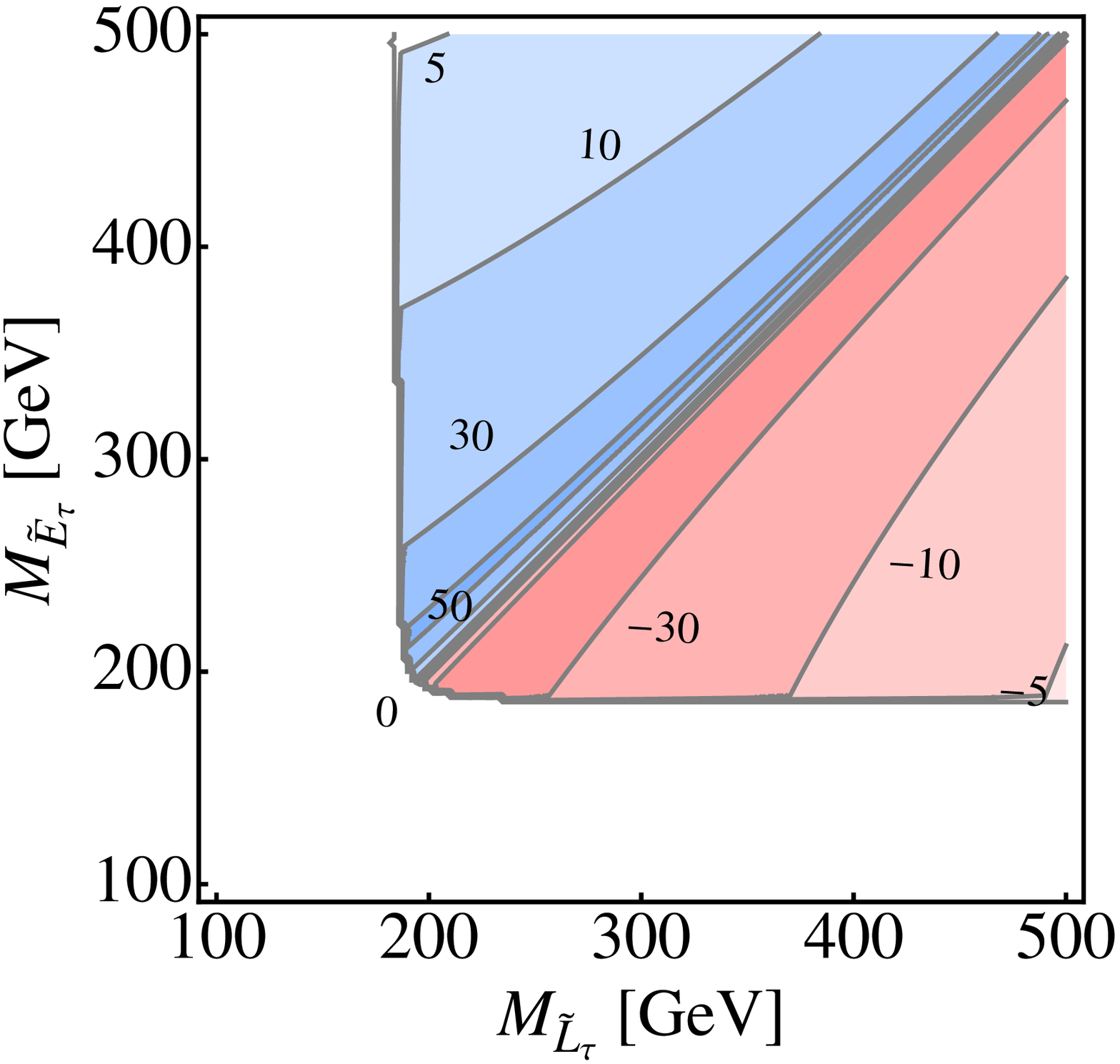}
\hfill
\label{fig:ML-MR_asymm_k1-plus-k2}}
\subfigure[]{  
  \includegraphics[width=0.95\columnwidth]{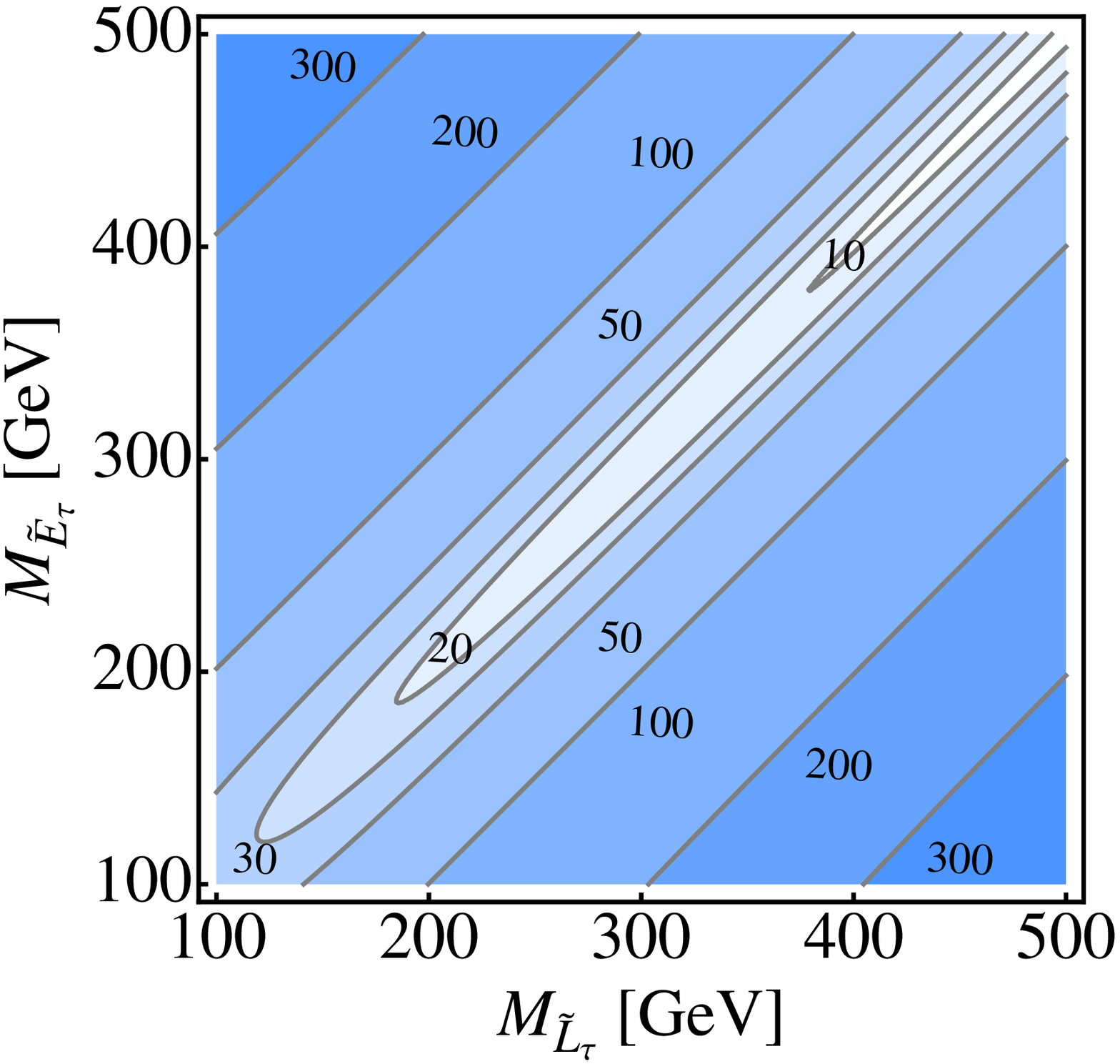}
\hfill  
\label{fig:ML-MR_Mstau2-Mstau1}}
  \hfill  
 \caption{%
     Contour lines of
     (a)~the sum of the $\tau$ polarization asymmetries 
     $\ACP$, Eq.~(\ref{eq:asym}), in percent, for the decays
     $\tilde\tau_1 \to \tau  \tilde\chi_2^0$ and 
     $\tilde\tau_2 \to \tau  \tilde\chi_2^0$,
     each in the stau rest frame and followed by
     $\tilde\chi_2^0 \to \ell_1^+ \tilde\ell_R^-$, 
     $\tilde\ell_R^- \to \tilde\chi^0_1 \ell_2^-$,
     for $\ell= e$ or $\mu$, see Fig.~\ref{Feyn}, and
      (b)~the stau mass splitting $m_{\stau_2} - m_{\stau_1}$ in GeV.
      Both plots are shown in the plane of the soft breaking parameters 
      of the stau sector,
     $M_{{\tilde{E}}_\tau}$--$M_{{\tilde{L}}_\tau}$, see
      Eqs.~(\ref{eq:left_stau}), (\ref{eq:right_stau}).
     The other MSSM parameters are given in Table~\ref{tab:benchmark.scenario}.
} 
\end{figure}

Note that also the stau branching ratios 
are similar in size; for example in our benchmark scenario we have
${\rm BR}(\tilde\tau_1\to\tau\tilde\chi_{2}^0)=18\%$, 
and
${\rm BR}(\tilde\tau_2\to\tau\tilde\chi_{2}^0)=30\%$. 
For the 
 $M_{{\tilde{E}}_\tau}$--$M_{{\tilde{L}}_\tau}$
plane shown in Fig.~\ref{fig:ML_MR_}, the decay branching ratio
${\rm BR}(\tilde\tau_1\to\tau\tilde\chi_{2}^0)$
is at least $10\%$, and that of $\stau_2$
is larger by roughly a factor of $2$ to $4$.
\clearpage

\section{Summary and conclusions}\label{sec:summary}
We have analyzed the normal tau polarization and the corresponding 
CP asymmetry in the two-body decay chain of a stau 
\begin{equation}
\stau_1\to \tau + {\tilde\chi}_2^0.
\end{equation}
The CP-sensitive parts appear only in the  spin-spin correlations,
which can be probed by the subsequent neutralino decay
\begin{equation}
{\tilde\chi}_2^0 \to \ell_1 + {\tilde\ell}_R;\quad 
{\tilde\ell}_R \to {\tilde\chi}_1^0 + \ell_2,
\end{equation}
for $\ell= e,\mu$.
The T-odd tau polarization normal to the plane spanned by the
$\tau$ and $\ell_1$ momenta, can then be used to define
a CP-odd tau polarization asymmetry.
It is based on a triple product, which
probes the CP phases of the trilinear scalar coupling parameter
$A_\tau$, the higgsino mass parameter $\mu$, and the U(1) gaugino mass
parameter $M_1$. \\ 
 
We have analyzed the analytical and numerical dependence
of the asymmetry on these parameters in detail. In particular,
for nearly degenerate staus where  the stau mixing is strong, 
the asymmetry obtains its maxima and can be larger than $70\%$. 
The normal tau polarization can thus be considered as an ideal CP observable
to probe the CP phases in the stau and neutralino sector of the MSSM.\\

Since the CP-sensitive parts appear only in the subsequent stau decay products 
the stau production process can be separated. Thus  both, ILC, and
LHC collider studies are possible. 
Concerning the kinematical dependence, the asymmetry is not Lorentz invariant, 
since it is based on a triple product. At the LHC,
staus are produced with a distinct boost distribution. Evaluated in the
laboratory frame, the resulting tau polarization asymmetries get
typically reduced  by a factor of two thirds,  compared to the stau
rest frame.   \\

We want to stress that a thorough experimental analysis, addressing background
processes, detector properties, and event rate reconstruction efficiencies,
will be needed in order to explore the measurability of CP phases in
the stau sector at the LHC or ILC. We hope that our work motivates
such a study.

\section*{Acknowledgements}\label{sec:acknowledge}
We thank M.~Drees and F.~von~der~Pahlen for enlightening discussions
and helpful comments. This work has been supported by MICINN project
FPA.2006-05294. AM was supported by the \textsl{Konrad Adenauer Stiftung}, 
BCGS, and a fellowship of Bonn University. HD was supported by the Hemholtz
Alliance ``Physics at the Terascale''  and BMBF ``Verbundprojekt HEP-Theorie''
under the contract 0509PDE. SK was supported by BCGS. OK acknowledges
support from CPAN.  

\appendix

\section{Momenta and spin vectors}\label{sec:momenta}

For the stau decay $\tilde \tau_m \to \tau\tilde\chi_i^0 $,
we choose the coordinate frame in the laboratory  (lab) system, 
such that the momentum of decaying $\stau$ points in the $z$-direction. 
 \begin{eqnarray}
  p^{\mu}_{\stau} &=& (E_{\stau}, 0, 0, |{\mathbf p}_{\stau}|), \\[2mm]
  p^{\mu}_{\tau} &=&  E_\tau
                     (1, \sin\theta_{\tau}, 0, \cos\theta_{\tau}), 
\end{eqnarray}
with the decay angle
$\theta_\tau =\varangle ({\mathbf p}_{\stau},{\mathbf p}_\tau)$, and
 \begin{eqnarray}
E_\tau &\approx & |{\mathbf p}_{\tau}|  \approx
               \frac{ (m_{\stau}^2 - m_{\schi}^2)} 
                    { 2( E_{\stau} - |{\mathbf p}_{\stau}| \cos\theta_{\tau}
               )}, 
\end{eqnarray}
in the limit $m_\tau \to 0$.
The momenta of the leptons from the subsequent neutralino decay
$\tilde\chi_i^0  \to \ell_1 \tilde\ell$; 
$\tilde\ell\to \tilde\chi_1^0 \ell_2$~(\ref{eq:staudec}),
can be parameterized by
\begin{eqnarray}
p^{\mu}_{\ell_1} &=& E_{\ell_1}
              (1, \sin\theta_1\cos\phi_1,
              \sin\theta_1\sin\phi_1,\cos\theta_1),\\[2mm]  
p^{\mu}_{\ell_2} &=& E_{\ell_2}
              (1, \sin\theta_2\cos\phi_2, \sin\theta_2\sin\phi_2,\cos\theta_2),
\end{eqnarray}
with the energies
\begin{eqnarray}
 E_{\ell_1} &=& \frac{ m_{\schi}^2-m_{\tilde{\ell}}^2}
                      { 2(E_{\schi}- |{\mathbf p}_{\schi}|\cos\theta_{D_1})},
                      \\  
 E_{\ell_2} &=& \frac{ m_{\tilde{\ell}}^2-m_{\schi}^2}
                     { 2(E_{\tilde{\ell}} - |{\mathbf
                      p}_{\tilde{\ell}}|\cos\theta_{D_2})}, 
\end{eqnarray} 
and the decay angles 
$\theta_{D_1} =\varangle ({\mathbf p}_{\schi},{\mathbf p}_{\ell_1})$,
$\theta_{D_2} =\varangle ({\mathbf p}_{\tilde\ell},{\mathbf p}_{\ell_2})$,
that is,  
\begin{eqnarray}
 \cos\theta_{D_1} &=& 
   \frac{({\mathbf p}_{\stau}-{\mathbf p}_{\tau})\cdot\hat{\mathbf p}_{\ell_1}}
       { | {\mathbf p}_{\stau}-{\mathbf p}_{\tau} | },\\[2mm]
\cos\theta_{D_2} &=& 
   \frac{({\mathbf p}_{\stau}-{\mathbf p}_{\tau}-{\mathbf p}_{\ell_1}
   )\cdot\hat{\mathbf p}_{\ell_2}} 
       { | {\mathbf p}_{\stau}-{\mathbf p}_{\tau}-{\mathbf p}_{\ell_1} | },
\end{eqnarray}
with the unit momentum vector $\hat{\mathbf p} = {\mathbf p}/|{\mathbf p}|$. 
We define the tau  spin vectors by
\begin{eqnarray}
 s_{\tau}^{1,\mu} &=& 
   \left(0, \frac{ {\mathbf s}_{\tau}^2 \times {\mathbf s}_{\tau}^3}
                 {|{\mathbf s}_{\tau}^2 \times {\mathbf s}_{\tau}^3|}\right),
  \qquad s_{\tau}^{2,\mu} = 
\left(0, \frac{{\mathbf p}_{\ell_1} \times {\mathbf p}_{\tau}}
             {|{\mathbf p}_{\ell_1} \times {\mathbf p}_{\tau}|}\right)
\nonumber, \\[2mm]   
s_{\tau}^{3,\mu}&=&\frac{1}{m_{\tau}}  
                      \left(|{\bf{p}}_{\tau}|,  
                      \frac{E_{\tau}}{|{\bf{p}}_{\tau}|}  
                       {\bf{p}}_{\tau}\right).
\label{eq:spintau}
\end{eqnarray}
The  spin vectors $s^a_{\tau},$ $a=1,2,3,$ for the tau,
and $s^b_{\tilde\chi_i^0},$ $b=1,2,3,$  for the neutralino $\tilde\chi^0_i$, 
fulfil completeness relations
\begin{eqnarray}
\sum_{a}s_\tau^{a,\,\mu} s_\tau^{a,\,\nu} &=& -g^{\mu\nu}
      + \frac{p_\tau^\mu p_\tau^\nu}{m_\tau^2},
\label{eq:completeness1}
\\[2mm]
\sum_{b}s_{\tilde\chi_i^0}^{b,\,\mu} s_{\tilde\chi_i^0}^{b,\,\nu}
    & =& -g^{\mu\nu}
      + \frac{p_{\tilde\chi_i^0}^\mu p_{\tilde\chi_i^0}^\nu}{m_{\tilde\chi_i^0}^2},
\label{eq:completeness2}
\end{eqnarray}
and they form orthonormal sets
\begin{eqnarray}
          s_{\tau}^a \cdot s_{\tau}^c &=& -\delta^{ac}, \qquad 
          s_{\tau}^a \cdot \hat{p}_{\tau} = 0, \\ [2mm]
          s_{\schi}^b \cdot s_{\schi}^c &=& -\delta^{bc}, \qquad 
          s_{\schi}^b \cdot \hat{p}_{\schi} = 0,
\end{eqnarray}
with $\hat{p}^{\mu} = p^{\mu} / m $.
Note that the asymmetry $\ACP$, Eq.~(\ref{P2}),
does not depend on the explicit form
of the neutralino spin vectors, since they are summed
in the amplitude squared, see Eq.~(\ref{eq:product1}), 
using the completeness relation.

\section{Phase space}\label{sec:phasespace}

The Lorentz invariant phase-space element for the stau decay chain, see
Eqs.~\eqref{eq:staudec} - \eqref{eq:decaychain}, can be decomposed into 
two-body phase-space elements~\cite{Bycki}
\begin{eqnarray}
\dLips(s_{\stau};\,p_{\ell_1},p_{\ell_2},p_{\tilde\chi_1^0})=
\dfrac{1}{(2\pi)^2}\displaystyle
\dLips(s_{\stau};\,p_\tau,p_{{\schi}})\nonumber\\[2mm]
\times 
{\rm d}s_{\schi}\,
\dLips(s_{\schi};\,p_{\ell_1},p_{\tilde\ell})\,
{\rm d}s_{\tilde\ell}\,
\dLips(s_{\tilde\ell};\,p_{\ell_2},p_{{\tilde\chi}_1^0}).\quad
\end{eqnarray} 
The different contributions are
\begin{eqnarray}
\dLips(s_{\stau};\,p_\tau,p_{\schi})&=&
       \dfrac{1}{4\pi}\dfrac{|{\bf p}_{\tau}|^2}
        {m_{\stau}^2-m_{\schi}^2}\sin\theta_\tau\,{\rm d}\theta_\tau,\quad\\
\dLips(s_{\schi};\,p_{\ell_1},p_{\tilde\ell})&=&
     \dfrac{1}{2(2\pi)^2}\dfrac{|{\bf p}_{\ell_1}|^2}
    {m_{\schi}^2-m_{\tilde\ell}^2}{\rm d}\Omega_1,\\
\dLips(s_{\tilde\ell};\,p_{\ell_2},p_{{\tilde\chi}_1^0})&=&
    \dfrac{1}{2(2\pi)^2}\dfrac{|{\bf p}_{\ell_2}|^2}
      {m_{\tilde\ell}^2-m_{{\tilde\chi}_1^0}^2}{\rm d}\Omega_2,
\end{eqnarray}
with $s_j=p_j^2$ and ${\rm d}\Omega_j=\sin\theta_j\,{\rm d}\theta_j\,{\rm
d}\phi_j$.

\section{Density matrix formalism}\label{sec:coefficients}

The coefficients of the stau decay matrix, Eq.~\eqref{rhoD}, are
\begin{eqnarray}
{\rm D}&=&
  \frac{g^2}{2}\left(|a_{mi}^{\stau}|^2+|b_{mi}^{\stau}|^2\right)(p_{\schi}\cdot p_\tau)
    \nonumber\\[2mm]
   && -g^2 \Re\{a_{mi}^{\stau} \,(b_{mi}^{\stau})^\ast\} m_{\schi}m_\tau,
\label{eq:D}\\
\Sigma_{\rm D}^a &=& 
        \,^{\;\,-}_{(+)}
        \frac{g^2}{2}\left(|a^{\stau}_{mi}|^2-|b^{\stau}_{mi}|^2\right)
        m_\tau(p_{\tilde\chi_i^0}\cdot s^a_{\tau}),
\label{eq:sigmaaD}\\
\Sigma_{\rm D}^b &=&
   \,^{\;\,-}_{(+)}
     \frac{g^2}{2} \left(|a^{\stau}_{mi}|^2-|b^{\stau}_{mi}|^2\right)
      m_{\schi} (p_\tau\cdot s_{\schi}^b),
\label{eq:sigmabD} \\
\Sigma_{\rm D}^{ab} &=& 
 \frac{g^2}{2}\left(|a^{\stau}_{mi}|^2 + |b^{\stau}_{mi}|^2\right)
         (s^a_{\tau}\cdot s^b_{{\tilde\chi_i^0}})m_\tau m_{\tilde\chi_i^0}
        \nonumber \\ [2mm] &&
         + g^2 \Re \{ a^{\stau}_{mi} (b^{\stau}_{mi})^\ast \}\times
 \nonumber \\ [2mm] &&
\left[  (s^a_{\tau}\cdot   p_{\tilde\chi_i^0}) (s^b_{\tilde\chi_i^0}\cdot
 p_\tau) 
       -(s^a_{\tau}\cdot s^b_{\tilde\chi_i^0}) (p_{\tilde\chi_i^0}  \cdot
 p_\tau) 
\right]
        \nonumber \\ [2mm] &&
        -g^2 \Im \{ a^{\stau}_{mi} (b^{\stau}_{mi})^\ast \}
         [s^{a}_{\tau},~ p_\tau,~ s^{b}_{\tilde\chi_i^0},~p_{\tilde\chi_i^0}].
\label{eq:sigmaabD} 
\end{eqnarray}
The formulas are given for the decay of a negatively charged stau,
$\stau_m \to \tau^- \tilde\chi_i^0 $. The signs in parentheses hold for 
the charge conjugated decay
$\stau_m^{\ast} \to \tau^+\tilde\chi_i^0 $.\\

Note that the terms proportional to $m_\tau$ in
Eqs.~\eqref{eq:D}, \eqref{eq:sigmaaD}, and \eqref{eq:sigmaabD}, are negligible
at high  
particle energies $E \gg m_\tau$, in particular  $\Sigma_{\rm D}^a$ 
can be neglected.\\

The coefficients of the $\tilde\chi_1^0$ decay matrix, Eq.~\eqref{rhoD1}, 
are~\cite{Kittel:2004rp}
\begin{eqnarray}
{\rm D}_1&=&
\frac{g^2}{2}|f_{\ell i}^R|^2(m_{\tilde\chi_i^0}^2 - m_{\tilde\ell}^2),
\label{eq:D1}\\[2mm]
\Sigma_{{\rm D}_1}^b&=&
\,^{\;\,+}_{(-)}
g^2|f_{\ell i}^R|^2 m_{\schi}(s_{\schi}^b\cdot p_{\ell_1}),
\label{eq:SigmabD1}
\end{eqnarray}
and the selectron decay factor is
\begin{eqnarray}
{\rm D}_2&=&g^2|f_{\ell_1}^R|^2(m_{\tilde\ell}^2-m_{\chi_1^0}^2).
\label{eq:D2}
\end{eqnarray}
The signs in parentheses hold for the charge conjugated processes, 
that is
$\tilde\chi^0_i \to \ell_1^- \tilde \ell_R^+$
in Eq.~(\ref{eq:SigmabD1}).\\

For the decay into a left slepton
 $\tilde\chi^0_i \to \ell_1^+ \tilde \ell_L^-$,
Eqs.~(\ref{eq:D1}), (\ref{eq:SigmabD1}), and (\ref{eq:D2}) 
read~\cite{Kittel:2004rp}
\begin{eqnarray}
{\rm D}_1 &=& \frac{g^2}{2} |f^{L}_{\ell i}|^2 (m_{\tilde\chi_i^0}^2
-m_{\tilde\ell}^2 ), 
\label{eq:D1selL} \\ [2mm]
\Sigma^b_{{\rm D}_1} &=& \,^{\;\,-}_{(+)} g^2 |f^{L}_{\ell i}|^2 
m_{\tilde\chi_i^0} (s^b_{\tilde\chi_i^0} \cdot p_{\ell_1}),
\label{eq:SigmabD1selL} \\ [2mm]
        {\rm D}_2&=& g^2 |f^{L}_{\ell 1}|^2 
        ( m_{\tilde\ell}^2-m_{\tilde\chi_1^0}^2 ),
\label{eq:D2selL}
\end{eqnarray}
respectively.
The expressions for Eqs.~(\ref{eq:product1}) and  (\ref{eq:product2})
have to be changed accordingly.
The sign in parenthesis in Eq.~(\ref{eq:SigmabD1selL}) holds
for the charge conjugated process
$\tilde\chi^0_i \to \ell_1^- \tilde \ell_L^+$.

\section{Stau decay widths}\label{sec:stauwidths}

The partial decay width for the decay $\tilde \tau_m \to \tau \tilde\chi_i^0$
in the stau rest frame is~\cite{Bartl:2002uy}
\begin{equation}
\Gamma(\tilde\tau_m \to \tau \tilde\chi_i^0)=
\frac{ m^2_{\tilde\tau} -m^2_{\tilde\chi^0_i}}
{4 \pi m^3_{\tilde\tau}}\,D,
\label{eq:widthstauchi0}
\end{equation}
with the decay function $D$ given in Eqs.~(\ref{eq:D}),
and the approximation $m_\tau= 0$.
For the decay $\tilde\tau_m \to \nu_\tau \tilde\chi_j^\pm$ 
the width is~\cite{Bartl:2002uy}
\begin{equation}
\Gamma(\tilde\tau_m \to \nu_\tau \tilde\chi_j^\pm)=
\frac{ (m^2_{\tilde\tau} -m^2_{\tilde\chi^\pm_j})^2 }
{16 \pi m^3_{\tilde\tau}}g^2
|l^{\tilde\tau}_{mj}|^2,
\label{eq:widthstauchar}
\end{equation}
with the stau-chargino-neutrino coupling~\cite{Haber:1984rc,Bartl:2002uy}
\begin{equation}
l_{mj}^{\tilde\tau} = -({\mathcal R}^{\tilde\tau}_{m1})^\ast \, U_{j1}+
              Y_\tau \,({\mathcal R}^{\tilde\tau}_{m2})^\ast \, U_{j2}, 
\label{eq:lstau}
\end{equation}
and the stau diagonalization matrix ${\mathcal R}^{\tilde\tau}$,  
Eq.~(\ref{eq:rstau}), the Yukawa coupling $Y_\tau$,  Eq.~(\ref{eq:yt}),
and the matrix $U$, that diagonalizes the chargino matrix~\cite{Haber:1984rc}, 
\begin{equation} 
    U^{\ast} \cdot {\mathcal M}_{\tilde\chi^\pm} \cdot V^{\dagger} = 
{\rm diag}(m_{\tilde\chi^\pm_{1}},m_{\tilde\chi^\pm_{2}}).
\label{wq:diagchar}   
\end{equation} 
The stau decay width for the entire decay chain, Eqs.~\eqref{eq:staudec} -
\eqref{eq:decaychain}, is then given by
\begin{eqnarray}
\lefteqn{ 
\Gamma(\tilde\tau\to\tau\ell_1\ell_2{\tilde\chi}_1^0)=
\nonumber
}
\\[2mm]
&=&
\frac{1}{2m_{\tilde\tau}}
\int\,|\mathcal{M}|^2\,
\dLips(s_{\stau};\,p_\tau,p_{\ell_1},p_{\ell_2},p_{\tilde\chi_1^0})
\\[2mm] &=&
\Gamma(\tilde\tau)\times {\rm BR}(\tilde\tau\to\tau \tilde\chi_i^0 )
                  \times {\rm BR}(\tilde\chi_i^0\to \ell_1\tilde\ell)
\nonumber\\
&&
                  \times {\rm BR}(\tilde\ell\to \ell_2\tilde\chi_1^0),
  \\
\end{eqnarray}
with the phase-space element $\dLips$, as given in the
Appendix~\ref{sec:momenta}, 
the amplitude squared
\begin{eqnarray}
|\mathcal{M}|^2 &=& 4|\Delta(\schi)|^2 |\Delta(\tilde\ell)|^2\, D\, D_1\, D_2, 
\end{eqnarray} 
obtained from  Eqs.~\eqref{eq:nosum} by summing the tau helicities
$\lambda_\tau$, $\lambda_\tau^\prime$.
The neutralino branching ratios are given, for example, 
in Ref.~\cite{Kittel:2004rp}, and we assume
${\rm BR}(\tilde\ell\to \ell_2\tilde\chi_1^0)=1$.
 We use the narrow width approximation for the propagators
\begin{equation}
\int |\Delta(j)|^2\,{\rm d}s_j = \dfrac{\pi}{m_j\Gamma_j},\label{NWA}
\end{equation}
which is justified for $\Gamma_j/m_j\ll 1$, which holds in our case with
$\Gamma_j\lesssim \mathcal{O}(1~{\rm GeV})$. Note, however, that in principle 
the naive $\mathcal{O}(\Gamma/m)$-expectation of the error can easily receive 
large off-shell corrections of an order of magnitude, and more, in particular 
at threshold, or due to interferences with other resonant, or non-resonant
processes~\cite{narrowwidth}.\\

  

\begin{thebibliography}{0}  

\bibitem{Belle}
%
  BELLE Collab., 
  A.~Abashian {\it et al.}, 
  Phys.\ Rev.\ Lett. {\bf 86}, 2509 (2001),
  hep-ex/0102018;\\
%
  BABAR Collab.,
  B.~Aubert {\it et al.},
  Phys.\ Rev.\ Lett. {\bf 86}, 2515 (2001),
  hep-ex/0102030.

\bibitem{Sakharov}
%
  A.~D.~Sakharov, Zh.\ Eksp.\ Teor.\ Fiz.\ Pis'ma {\bf 5}, 32 (1967),
  JETP\ Lett. {\bf 91B}, 24 (1967).

\bibitem{Haber:1984rc} 
  H.~E.~Haber and G.~L.~Kane, 
  Phys.\ Rept.\  {\bf 117} (1985) 75;\\
%
  H.~P.~Nilles, 
  Phys.\ Rept.\  {\bf 110} (1984) 1;\\
%
  M.~Drees, R.~Godbole and P.~Roy, 
  ``Theory and phenomenology of sparticles: An account of four-dimensional N=1 
  supersymmetry in high energy physics,'' 
  {\it  Hackensack, USA: World Scientific (2004)}.
%


\bibitem{Li:2010ax}
  For a recent review of the CP-violation constraints within the MSSM, see\\
  Y.~Li, S.~Profumo and M.~Ramsey-Musolf,
  JHEP {\bf 1008}, 062 (2010)
  [arXiv:1006.1440 [hep-ph]].


%
\bibitem{cancellations1} 
%
  T.~Ibrahim and P.~Nath, 
  Phys.\ Rev.\  D {\bf 57} (1998) 478 
  [Erratum-ibid.\  D {\bf 58} (1998\ ERRAT,D60,079903.1999\  
  ERRAT,D60,119901.1999) 019901] 
  [arXiv:hep-ph/9708456]; 
%
  Phys.\ Lett.\  B {\bf 418} (1998) 98 
  [arXiv:hep-ph/9707409]; 
%
  [Erratum-ibid.\  D {\bf 60} (1999) 099902] 
  [arXiv:hep-ph/9807501]; 
%
  Phys.\ Rev.\  D {\bf 61} (2000) 093004 
  [arXiv:hep-ph/9910553];\\ 
%
  M.~Brhlik, G.~J.~Good and G.~L.~Kane, 
  Phys.\ Rev.\  D {\bf 59} (1999) 115004 
  [arXiv:hep-ph/9810457];\\ 
%
  S.~Yaser Ayazi and Y.~Farzan, 
  Phys.\ Rev.\  D {\bf 74}, 055008 (2006) 
  [arXiv:hep-ph/0605272].
%
%
\bibitem{cancellations2} 
%
see, e.g., 
  A.~Bartl, T.~Gajdosik, W.~Porod, P.~Stockinger and H.~Stremnitzer, 
  Phys.\ Rev.\  D {\bf 60} (1999) 073003 
  [arXiv:hep-ph/9903402];\\ 
%
  A.~Bartl, T.~Gajdosik, E.~Lunghi, A.~Masiero, W.~Porod, H.~Stremnitzer and  
  O.~Vives,  
  Phys.\ Rev.\  D {\bf 64} (2001) 076009 
  [arXiv:hep-ph/0103324];\\ 
%
  L.~Mercolli and C.~Smith, 
  Nucl.\ Phys.\  B {\bf 817}, 1 (2009) 
  [arXiv:0902.1949 [hep-ph]];\\ 
%
  V.~D.~Barger, T.~Falk, T.~Han, J.~Jiang, T.~Li and T.~Plehn, 
  Phys.\ Rev.\  D {\bf 64} (2001) 056007 
  [arXiv:hep-ph/0101106]. 
%
\bibitem{cancellations3} 
%
  A.~Bartl, W.~Majerotto, W.~Porod and D.~Wyler, 
  Phys.\ Rev.\  D {\bf 68} (2003) 053005 
  [arXiv:hep-ph/0306050];\\ 
%
  K.~A.~Olive, M.~Pospelov, A.~Ritz and Y.~Santoso, 
  Phys.\ Rev.\  D {\bf 72} (2005) 075001 
  [arXiv:hep-ph/0506106]; \\
%
  J.~R.~Ellis, J.~S.~Lee and A.~Pilaftsis, 
  JHEP {\bf 0810} (2008) 049  
  [arXiv:0808.1819 [hep-ph]];\\
%
  S.~A.~Abel, A.~Dedes and H.~K.~Dreiner,
  JHEP {\bf 0005} (2000) 013
  [arXiv:hep-ph/9912429].




\bibitem{totalsigma}
%
  S.~Y.~Choi, A.~Djouadi, M.~Guchait, J.~Kalinowski, H.~S.~Song and 
  P.~M.~Zerwas,  
  Eur.\ Phys.\ J.\  C {\bf 14} (2000) 535 
  [arXiv:hep-ph/0002033];\\ 
%
  J.~L.~Kneur and G.~Moultaka,
  Phys.\ Rev.\ D {\bf 61} (2000) 095003,
  arXiv:hep-ph/9907360.
%
\bibitem{masses}
%
  A.~Pilaftsis {\it et al.},
  Phys.\ Lett.\ B {\bf 435} (1998) 88,
  arXiv:hep-ph/9805373;\\
%
  D.~A.~Demir {\it et al.},
  Phys.\ Rev.\ D {\bf 60} (1999) 055006,
  arXiv:hep-ph/9807336;\\
%
  A.~Pilaftsis and C.~E.~M.~Wagner,
  Nucl.\ Phys.\ B {\bf 553} (1999) 3,
  arXiv:hep-ph/9902371;\\
%
  S.~Y.~Choi, M.~Drees, and J.~S.~Lee,
  Phys.\ Lett.\ B {\bf 481} (2000) 57,
  arXiv:hep-ph/0002287;\\
%
  J.~L.~Kneur and G.~Moultaka,
  Phys.\ Rev.\ D {\bf 59} (1999) 015005,
  arXiv:hep-ph/9807336.
%
\bibitem{BRs}
%
  A.~Bartl, S.~Hesselbach, K.~Hidaka, T.~Kernreiter and W.~Porod, 
  Phys.\ Rev.\  D {\bf 70} (2004) 035003  
  [arXiv:hep-ph/0311338];\\ 
%
  K.~Rolbiecki, J.~Tattersall and G.~Moortgat-Pick, 
  arXiv:0909.3196 [hep-ph]. 

%
\bibitem{rateasymBRs}
%
  H.~Eberl, T.~Gajdosik, W.~Majerotto and B.~Schrausser, 
  Phys.\ Lett.\  B {\bf 618} (2005) 171 
  [arXiv:hep-ph/0502112];\\ 
%
  E.~Chistova, H.~Eberl, W.~Majerotto, and S.~Kraml,
  Nucl.\ Phys.\ B {\bf 639} (2002) 263,
  arXiv:hep-ph/0205227;\\
%
  E.~Chistova, H.~Eberl, W.~Majerotto, and S.~Kraml,
  JHEP {\bf 12} (2002) 021,
  arXiv:hep-ph/0211063;\\
%
  M.~Frank and I.~Turan,
  Phys.\ Rev.\ D {\bf 76} (2007) 016001,
  arXiv:hep-ph/0703184;\\
%
  M.~Frank and I.~Turan, 
  Phys.\ Rev.\  D {\bf 76} (2007) 076008 
  [arXiv:0708.0026 [hep-ph]]. 
%
\bibitem{rateasymsigma}
%
  E.~Christova, H.~Eberl, E.~Ginina and W.~Majerotto,
  Phys.\ Rev.\  D {\bf 79}, 096005 (2009)
  [arXiv:0812.4392 [hep-ph]].
%
\bibitem{angulardistrib}
%
  E.~Christova, H.~Eberl, E.~Ginina and W.~Majerotto, 
  JHEP {\bf 0702} (2007) 075  
  [arXiv:hep-ph/0612088]. 
%



\bibitem{res}
  A.~Pilaftsis,
  Nucl.\ Phys.\  B {\bf 504}, 61 (1997)
  [arXiv:hep-ph/9702393];\\
%
  S.~Y.~Choi, J.~Kalinowski, Y.~Liao and P.~M.~Zerwas,
  Eur.\ Phys.\ J.\  C {\bf 40}, 555 (2005)
  [arXiv:hep-ph/0407347];\\
%
  H.~K.~Dreiner, O.~Kittel and F.~von der Pahlen,
  JHEP {\bf 0801}, 017 (2008)
  [arXiv:0711.2253 [hep-ph]];\\
%
  O.~Kittel and F.~von der Pahlen,
  JHEP {\bf 0808}, 030 (2008)
  [arXiv:0806.4534 [hep-ph]];\\
%
  M.~Nagashima, K.~Kiers, A.~Szynkman, D.~London, J.~Hanchey and K.~Little,
  Phys.\ Rev.\  D {\bf 80}, 095012 (2009)
  [arXiv:0907.1063 [hep-ph]].

\bibitem{tripleproducts}
For studies with neutralino 3-body decays at the ILC, see\\ 
  Y.~Kizukuri and N.~Oshimo, 
  Phys.\ Lett.\  B {\bf 249} (1990) 449;\\ 
%
  S.~Y.~Choi, H.~S.~Song and W.~Y.~Song, 
  Phys.\ Rev.\  D {\bf 61} (2000) 075004  
  [arXiv:hep-ph/9907474];\\
%
  For further studies with neutralino 2-body and 3-body decays at the ILC,
  see\\  
  A.~Bartl, H.~Fraas, O.~Kittel and W.~Majerotto, 
  Phys.\ Rev.\  D {\bf 69} (2004) 035007  
  [arXiv:hep-ph/0308141];\\ 
%
  A.~Bartl, H.~Fraas, O.~Kittel and W.~Majerotto, 
  Eur.\ Phys.\ J.\  C {\bf 36} (2004) 233  
  [arXiv:hep-ph/0402016];\\ 
%
  J.~A.~Aguilar-Saavedra, 
  Nucl.\ Phys.\  B {\bf 697} (2004) 207  
  [arXiv:hep-ph/0404104];\\ 
%
  A.~Bartl, H.~Fraas, S.~Hesselbach, K.~Hohenwarter-Sodek and 
  G.~A.~Moortgat-Pick,  
  JHEP {\bf 0408} (2004) 038 
  [arXiv:hep-ph/0406190];\\ 
%
  S.~Y.~Choi, B.~C.~Chung, J.~Kalinowski, Y.~G.~Kim and K.~Rolbiecki, 
  Eur.\ Phys.\ J.\  C {\bf 46} (2006) 511 
  [arXiv:hep-ph/0504122];\\ 
%
  For studies with chargino 2-body decays at the ILC, see\\ 
  A.~Bartl, H.~Fraas, O.~Kittel and W.~Majerotto, 
  Phys.\ Lett.\  B {\bf 598} (2004) 76  
  [arXiv:hep-ph/0406309];\\ 
%
  O.~Kittel, A.~Bartl, H.~Fraas and W.~Majerotto, 
  Phys.\ Rev.\  D {\bf 70} (2004) 115005  
  [arXiv:hep-ph/0410054];\\ 
%
For studies with chargino 3-body decays at the ILC, see\\ 
  Y.~Kizukuri and N.~Oshimo, 
  arXiv:hep-ph/9310224;\\ 
%
  A.~Bartl, H.~Fraas, S.~Hesselbach, K.~Hohenwarter-Sodek, T.~Kernreiter and 
  G.~Moortgat-Pick,  
  Eur.\ Phys.\ J.\  C {\bf 51} (2007) 149  
  [arXiv:hep-ph/0608065]. 
%

\bibitem{CPreview} 
For recent reviews see, for example,\\ 
  G.~Moortgat-Pick, K.~Rolbiecki, J.~Tattersall and P.~Wienemann, 
  arXiv:0910.1371 [hep-ph];\\ 
%
  O.~Kittel, 
  arXiv:0904.3241 [hep-ph].
%

\bibitem{LHC}
%
  S.~Abdullin {\it et al.}  [CMS Collaboration], 
  J.\ Phys.\ G {\bf 28} (2002) 469  
  [arXiv:hep-ph/9806366];\\ 
%
 ATLAS collab.,   
  {\it ATLAS detector and physics performance. Technical design report.
  Vol. 2},                                                            
  CERN-LHCC-99-15;\\ 
%
  G.~Weiglein {\it et al.}  [LHC/LC Study Group], 
  arXiv:hep-ph/0410364. 

\bibitem{ILC}
%
 J.~Brau {\it et al.}  [ILC Collaboration], 
  arXiv:0712.1950 [physics.acc-ph];\\ 
%
  J.~A.~Aguilar-Saavedra {\it et al.}  [ECFA/DESY LC Physics Working Group], 
  arXiv:hep-ph/0106315;\\ 
%
  T.~Abe {\it et al.}  [American Linear Collider Working Group], 
  arXiv:hep-ex/0106055;\\ 
%
  K.~Abe {\it et al.}  [ACFA Linear Collider Working Group], 
  arXiv:hep-ph/0109166;\\ 
%
  J.~A.~Aguilar-Saavedra {\it et al.}, 
  Eur.\ Phys.\ J.\ C {\bf 46} (2006) 43 
  [arXiv:hep-ph/0511344]. 
%


\bibitem{Semertzidis:2004uu} 
  Y.~K.~Semertzidis, 
  Nucl.\ Phys.\ Proc.\ Suppl.\  {\bf 131} (2004) 244  
  [arXiv:hep-ex/0401016]; \\
%
  J.~R.~Ellis, S.~Ferrara and D.~V.~Nanopoulos, 
  Phys.\ Lett.\  B {\bf 114} (1982) 231;\\ 
%
  W.~Buchmuller and D.~Wyler, 
  Phys.\ Lett.\  B {\bf 121} (1983) 321;\\ 
%
  F.~del Aguila, M.~B.~Gavela, J.~A.~Grifols and A.~Mendez, 
  Phys.\ Lett.\  B {\bf 126} (1983) 71 
  [Erratum-ibid.\  B {\bf 129} (1983) 473];\\ 
%
  D.~V.~Nanopoulos and M.~Srednicki, 
  Phys.\ Lett.\  B {\bf 128} (1983) 61;\\ 
%
  M.~Dugan, B.~Grinstein and L.~J.~Hall, 
  Nucl.\ Phys.\  B {\bf 255} (1985) 413;\\ 
%
  C.~S.~Huang and W.~Liao, 
  Phys.\ Rev.\  D {\bf 62} (2000) 016008 
  [arXiv:hep-ph/0001174]. 

\bibitem{Choi:2004rf} 
  S.~Y.~Choi, M.~Drees and B.~Gaissmaier, 
  Phys.\ Rev.\  D {\bf 70} (2004) 014010 
  [arXiv:hep-ph/0403054]. 

\bibitem{Deppisch:2009nj}
  F.~Deppisch and O.~Kittel,
  JHEP {\bf 0909}, 110 (2009)
  [Erratum-ibid.\  {\bf 1003}, 091 (2010)]
  [arXiv:0905.3088 [hep-ph]].

\bibitem{MoortgatPick:2009jy}
  G.~Moortgat-Pick, K.~Rolbiecki, J.~Tattersall and P.~Wienemann,
  JHEP {\bf 1001}, 004 (2010)
  [arXiv:0908.2631 [hep-ph]];\\
%
  G.~Moortgat-Pick, K.~Rolbiecki and J.~Tattersall,
  arXiv:1008.2206 [hep-ph].

\bibitem{Bartl:2004jr} 
  A.~Bartl, E.~Christova, K.~Hohenwarter-Sodek and T.~Kernreiter, 
  Phys.\ Rev.\  D {\bf 70} (2004) 095007 
  [arXiv:hep-ph/0409060].
  
\bibitem{Ellis:2008hq} 
  J.~Ellis, F.~Moortgat, G.~Moortgat-Pick, J.~M.~Smillie and J.~Tattersall, 
  Eur.\ Phys.\ J.\  C {\bf 60}, 633 (2009) 
  [arXiv:0809.1607 [hep-ph]]. 

\bibitem{Bartl:2006hh} 
  A.~Bartl, E.~Christova, K.~Hohenwarter-Sodek and T.~Kernreiter, 
  JHEP {\bf 0611} (2006) 076  
  [arXiv:hep-ph/0610234]. 

\bibitem{Deppisch:2010nc}
  F.~F.~Deppisch and O.~Kittel,
  JHEP {\bf 1006}, 067 (2010)
  [arXiv:1003.5186 [hep-ph]].

\bibitem{Bartl:2003ck} 
  A.~Bartl, H.~Fraas, T.~Kernreiter and O.~Kittel, 
  Eur.\ Phys.\ J.\  C {\bf 33} (2004) 433 
  [arXiv:hep-ph/0306304]. 
%

\bibitem{Bartl:2003gr} 
  A.~Bartl, T.~Kernreiter and O.~Kittel, 
  Phys.\ Lett.\  B {\bf 578} (2004) 341  
  [arXiv:hep-ph/0309340];\\ 
%
  O.~Kittel, 
  arXiv:hep-ph/0311169. 

\bibitem{MKD}
  H.~Dreiner, O.~Kittel and A.~Marold,
arXiv:1001.4714 [hep-ph]
.
%

\bibitem{Haber:1994pe} 
  H.~E.~Haber,  
  Proceedings of the 21st SLAC Summer Institute on Particle Physics, 
  eds. L.~DeProcel, Ch.~Dunwoodie,  
  Stanford 1993, 231. 
  [arXiv:hep-ph/9405376]. 

\bibitem{Bartl:2002uy} 
  A.~Bartl, K.~Hidaka, T.~Kernreiter and W.~Porod, 
  Phys.\ Lett.\  B {\bf 538} (2002) 137  
%
  Phys.\ Rev.\  D {\bf 66} (2002) 115009 
  [arXiv:hep-ph/0207186].

\bibitem{Kittel:2004rp}  
  O.~Kittel,  
  [arXiv:hep-ph/0504183].  

\bibitem{Renard:1981de}  
  F.~M.~Renard, 
  ``Basics Of Electron Positron Collisions,''  
  {\it Dreux, France: Editions  Frontieres (1981)}.   

\bibitem{Luders:1954zz} 
  G.~Luders, 
  Kong.\ Dan.\ Vid.\ Sel.\ Mat.\ Fys.\ Med.\ {\bf 28N5} (1954) 1;\\ 
%
  W.~Pauli, 
  ``Niels Bohr and the development of physics,'' 
  {\it New York, USA: Mc Graw-Hill (1955)};\\ 
%
  R.~Jost, 
  Helv.\ Phys.\ Acta {\bf 30} (1957) 409;\\ 
%
  R.~Jost, 
  Helv.\ Phys.\ Acta {\bf 36} (1963) 77;\\ 
%
  R.~F.~Streater and A.~S.~Wightman, 
  ``PCT, spin and statistics, and all that,'' 
  {\it Redwood City, USA: Addison-Wesley (1989)  (Advanced book 
  classics).}  

\bibitem{Alwall:2007st}
  J.~Alwall {\it et al.},
  JHEP {\bf 0709}, 028 (2007)
  [arXiv:0706.2334 [hep-ph]].

\bibitem{Frere:1983}
  J.~M.~Fr{'e}re, D.~R.~T.~Jones, and S.~Rabi,
  Nucl.\ Phys.\ {\bf B222} (1983) 11;
%
  M.~Claudson, L.~J.~Hall, and I.~Hinchliffe,
  Nucl.\ Phys.\ {\bf B228} (1983) 501;
%
  C.~Kounnas, A.~B.~Lahanas, D.~V.~Nanopoulos, and M.~Quir{'o}s,
  Nucl.\ Phys.\ {\bf B236} (1984) 438;
%
  J.~F.~Gunion, H.~E.~Haber, and M.~Sher,
  Nucl.\ Phys.\ {\bf B306} (1988) 1.

\bibitem{Bycki}  
%
  E.~Byckling, K.~Kajantie,  
  ``Particle Kinematics,''  
  {\it London, England: John~Wiley\& Sons (1973)};\\ 
%
  G.~Costa {\it et al.},  
  TEPP ``Kinematics and Symmetries'', Bd.1, ed. M.~Nicoli\'{c},  
  {\it Paris, France: Institut national de physique nucl\'{e}aire et de 
  physique des particules (1979)}.  

%
\bibitem{narrowwidth}  
%
  K.~Hagiwara {\it et al.},  
  Phys.\ Rev.\ D {\bf 73} (2006) 055005  
  [arXiv:hep-ph/0512260];\\  
%
  D.~Berdine, N.~Kauer and D.~Rainwater, 
  Phys.\ Rev.\ Lett.\  {\bf 99} (2007) 111601 
  [arXiv:hep-ph/0703058;\\ 
%
  N.~Kauer, 
  Phys.\ Lett.\  B {\bf 649} (2007) 413 
  [arXiv:hep-ph/0703077]; 
%
  JHEP {\bf 0804} (2008) 055 
  [arXiv:0708.1161 [hep-ph]];\\ 
%
  C.~F.~Uhlemann and N.~Kauer, 
  Nucl.\ Phys.\  B {\bf 814} (2009) 195 
  [arXiv:0807.4112 [hep-ph]];\\ 
%
  M.~A.~Gigg and P.~Richardson, 
  arXiv:0805.3037 [hep-ph]. 

\end{thebibliography}
\end{document}